\documentclass[a4paper]{report}
\usepackage[english]{babel}
\usepackage{amstex}
\usepackage{cite}
\title{Renormalization Group Studies of Vertex Models}
\author{Saibal Mitra\\Supervisor: Prof.\ Dr.\ H.\ van Beijeren\\
ITF-UU-99/04}
\begin{document}
\maketitle
\def\haak#1{\left(#1\right)}
\def\rhaak#1{\left [#1\right]}
\def\lhaak#1{\left | #1\right |}
\def\ahaak#1{\left\{#1\right\}}
\def\gem#1{\left\langle #1\right\rangle}
\def\gemc#1#2{\left\langle\left\langle\left. #1\right | #2 \right\rangle\right\rangle }
\def\geml#1{\left\langle #1\right.}
\def\gemr#1{\left. #1\right\rangle}
\def\haakl#1{\left(#1\right.}
\def\haakr#1{\left.#1\right)}
\def\rhaakl#1{\left[#1\right.}
\def\rhaakr#1{\left.#1\right]}
\def\lhaakl#1{\left |#1\right.}
\def\lhaakr#1{\left.#1\right |}
\def\rge{renormalization group equations}
\def\Rge{Renormalization group equations}
\def\rgee{renormalization group equation}
\def\hee{h^{\haak{1}}}
\def\htw{h^{\haak{2}}}
\def\stfm{staggered F-model}
\def\half{\frac{1}{2}}
\def\tr{\operatorname{Tr}}
\tableofcontents
\chapter{Introduction}
In this thesis we use renormalization group methods to study the critical behaviour
of the \stfm. The \stfm, defined in chapter \ref{dstfm}, can be used as a model 
of a facet of a BCC crystal in the (100) direction. 

We first use known exact results to map the \stfm\ to a sine-Gordon type model
(defined in section \ref{eff}), and study the \rge\ for this model using momentum 
shell integration techniques.
The map to the sine-Gordon model is constructed using an exact result on the
long range part of the height-height correlation function of the F-model (i.e.\
the \stfm\ at zero staggered field) \cite{jeps}. The results we obtain are
the phase diagram of the \stfm\ and the leading singularity in the free energy.

To get more results, e.g.\ the next to leading singularity in the free energy,
we need a larger
source of information than is available in the form of the long range part of
the height-height correlation function. 
It turns out that the free fermion method, upon which Baxter's original solution 
is based, is flexible enough to admit a perturbative expansion about the free 
fermion line. By calculating the singular part of the free energy perturbatively
about the free fermion line, one can construct a map from the \stfm\
to the sine-Gordon model by demanding that it correctly reproduces the singular
part of the free energy. The idea thus is to use the mapping to the sine-Gordon type
model as an extrapolation technique. To make this approach practical we:
\begin{enumerate}
\item Develop in chapter \ref{dren} a simple diagrammatic method 
to find the \rge\ for a given sine-Gordon type model.
This method is based on a combination of functional Feynman rules and the
operator product expansion. 
\item Rewrite in section \ref{lce} the perturbative expansion about 
the free fermion line as a linked cluster expansion.
\end{enumerate}
Although these two results make it possible to construct a map to a sine-Gordon
type model in a systematic way, the actual construction of this map is beyond  
the scope of this thesis. We do, however, explicitly calculate the first order 
correction to Baxter's result. This allows us to verify results which previously
could only be obtained using renormalization group arguments. 
\section{Summary}
\begin{description}
\item In chapter \ref{dstfm} we introduce the \stfm\ and discuss the equivalence
with the BCSOS model. 
\item In chapter \ref{dren} a systematic method is derived to generate the \rge\
\item In chapter \ref{apren} we first discuss the application of the
renormalization group in the calculation of critical exponents. We then proceed
to obtain the phase diagram of the \stfm\ and
calculate the leading singularity in the 
free energy by using the information present in the form
of the asymptotic form of the height-height correlation function 
\item In chapter \ref{ffc} we first present a derivation of Baxter's 
exact solution. We then perturbatively lift the free fermion condition. This
allows one to write the free energy of the \stfm\ as a perturbative expansion
about the free fermion line. Baxter's exact solution can thus be seen as the
zeroth order term in this expansion. We explicitely calculate the first order
term. To facilitate the computation of the higher order terms we derive a linked
cluster method in section \ref{lce}.
\end{description}
\chapter{Definition of the \stfm\ }\label{dstfm}
In this chapter we shall introduce the six-vertex model, of which 
the \stfm\ is a special case. Some known results are discussed.
\section{The six-vertex model}
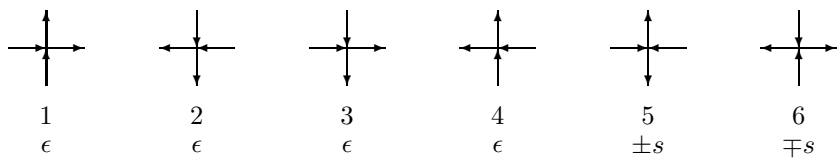
\begin{figure}
\setlength{\unitlength}{0.0165 \textwidth}
\begin{picture}(60,10)
\put(5,7.5){\vector(1,0){2.5}}
\put(2.5,7.5){\vector(1,0){2.5}}
\put(5,5){\vector(0,1){2.5}}
\put(5,7.5){\vector(0,1){2.5}}
\put(5,3){\makebox(0,0){1}}
\put(5,1){\makebox(0,0){$\epsilon $}}

\put(17.5,7.5){\vector(-1,0){2.5}}
\put(15,7.5){\vector(-1,0){2.5}}
\put(15,10){\vector(0,-1){2.5}}
\put(15,7.5){\vector(0,-1){2.5}}
\put(15,3){\makebox(0,0){2}}
\put(15,1){\makebox(0,0){$\epsilon $}}

\put(25,7.5){\vector(1,0){2.5}}
\put(22.5,7.5){\vector(1,0){2.5}}
\put(25,10){\vector(0,-1){2.5}}
\put(25,7.5){\vector(0,-1){2.5}}
\put(25,3){\makebox(0,0){3}}
\put(25,1){\makebox(0,0){$\epsilon $}}

\put(37.5,7.5){\vector(-1,0){2.5}}
\put(35,7.5){\vector(-1,0){2.5}}
\put(35,5){\vector(0,1){2.5}}
\put(35,7.5){\vector(0,1){2.5}}
\put(35,3){\makebox(0,0){4}}
\put(35,1){\makebox(0,0){$\epsilon $}}

\put(47.5,7.5){\vector(-1,0){2.5}}
\put(42.5,7.5){\vector(1,0){2.5}}
\put(45,7.5){\vector(0,1){2.5}}
\put(45,7.5){\vector(0,-1){2.5}}
\put(45,3){\makebox(0,0){5}}
\put(45,1){\makebox(0,0){$\pm s $}}

\put(55,7.5){\vector(-1,0){2.5}}
\put(55,7.5){\vector(1,0){2.5}}
\put(55,10){\vector(0,-1){2.5}}
\put(55,5){\vector(0,1){2.5}}
\put(55,3){\makebox(0,0){6}}
\put(55,1){\makebox(0,0){$\mp s $}}
\end{picture}
\setlength{\unitlength}{1 pt}
\caption{\small The six vertices and their energies. The upper and lower
signs correspond to the two sublattices.}\label{vrt}
\end{figure}

The six-vertex model can be defined as follows:
place arrows on the bonds of a square lattice so that there are two arrows
pointing into each vertex. Six types of vertices can arise (hence 
the name of the model). These vertices are shown in fig.\ \ref{vrt}.
By giving each vertex-type a (position-dependent) energy the model is defined.
These models were first introduced to study ferroelectric systems. Later it
was shown that six-vertex models can be mapped to solid-on-solid (SOS) models \cite{vby}.
Only a few of these models can be solved exactly. These include the
free fermion models \cite{fan,wu} and models that can can be solved using a 
(generalized) Bethe Ansatz \cite{bxt2,lieb1,lieb2,lieb3,bxt1}. To define
the \stfm, we divide the lattice into two sublattices A and B, such that the
nearest neighbor of an A vertex is a B vertex. The vertex energies are chosen
as indicated in fig.\ \ref{vrt}. When the the staggered field ($ s $) vanishes the model reduces  
to the F-model, which has been solved by Lieb \cite{lieb2}. 
For nonzero staggered field the model can be solved when $\beta\epsilon=\frac{1}{2}\ln\haak{2}$
\cite{bxt}. 
\section{Six-vertex models and SOS models}\label{sos}
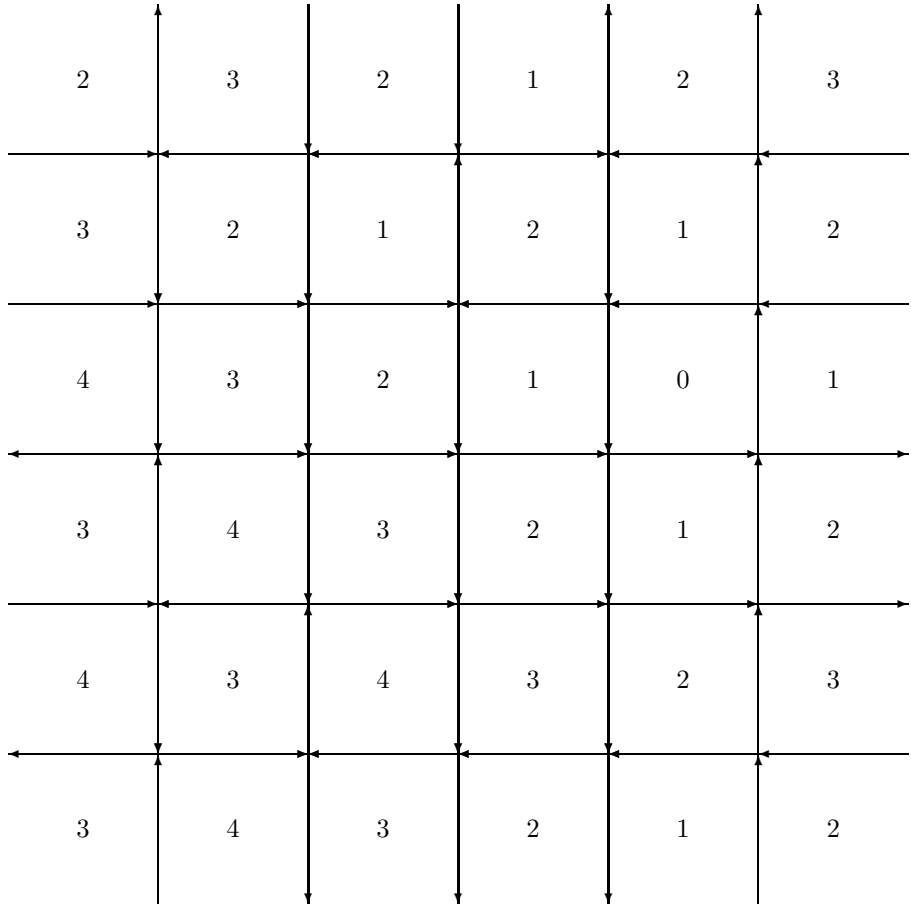
\begin{figure}
\setlength{\unitlength}{0.01645\textwidth}
\begin{picture}(60,60)
\newsavebox{\vrton}\newsavebox{\vrttw}\newsavebox{\vrtth}\newsavebox{\vrtfo}\newsavebox{\vrtfi}\newsavebox{\vrtsi}
\savebox{\vrton}(0,0){\begin{picture}(20,20)
\put(10,10){\vector(0,1){10}}
\put(10,0){\vector(0,1){10}}
\put(0,10){\vector(1,0){10}}
\put(10,10){\vector(1,0){10}}
\end{picture}}

\savebox{\vrttw}(0,0){\begin{picture}(20,20)
\put(10,20){\vector(0,-1){10}}
\put(10,10){\vector(0,-1){10}}
\put(20,10){\vector(-1,0){10}}
\put(10,10){\vector(-1,0){10}}
\end{picture}}

\savebox{\vrtth}(0,0){\begin{picture}(20,20)
\put(10,10){\vector(0,1){10}}
\put(10,0){\vector(0,1){10}}
\put(20,10){\vector(-1,0){10}}
\put(10,10){\vector(-1,0){10}}
\end{picture}}

\savebox{\vrtfo}(0,0){\begin{picture}(20,20)
\put(10,20){\vector(0,-1){10}}
\put(10,10){\vector(0,-1){10}}
\put(0,10){\vector(1,0){10}}
\put(10,10){\vector(1,0){10}}
\end{picture}}

\savebox{\vrtfi}(0,0){\begin{picture}(20,20)
\put(10,10){\vector(0,1){10}}
\put(10,10){\vector(0,-1){10}}
\put(0,10){\vector(1,0){10}}
\put(20,10){\vector(-1,0){10}}
\end{picture}}

\savebox{\vrtsi}(0,0){\begin{picture}(20,20)
\put(10,20){\vector(0,-1){10}}
\put(10,0){\vector(0,1){10}}
\put(10,10){\vector(-1,0){10}}
\put(10,10){\vector(1,0){10}}
\end{picture}}

\put(10,10){\usebox{\vrtsi}}
\put(10,20){\usebox{\vrtfi}}
\put(10,30){\usebox{\vrtsi}}
\put(10,40){\usebox{\vrtfo}}
\put(10,50){\usebox{\vrtfi}}
\put(20,10){\usebox{\vrtfi}}
\put(20,20){\usebox{\vrtsi}}
\put(20,30){\usebox{\vrtfo}}
\put(20,40){\usebox{\vrtfo}}
\put(20,50){\usebox{\vrttw}}
\put(30,10){\usebox{\vrttw}}
\put(30,20){\usebox{\vrtfo}}
\put(30,30){\usebox{\vrtfo}}
\put(30,40){\usebox{\vrtfi}}
\put(30,50){\usebox{\vrtsi}}
\put(40,10){\usebox{\vrttw}}
\put(40,20){\usebox{\vrtfo}}
\put(40,30){\usebox{\vrtfo}}
\put(40,40){\usebox{\vrttw}}
\put(40,50){\usebox{\vrtfi}}
\put(50,10){\usebox{\vrtth}}
\put(50,20){\usebox{\vrton}}
\put(50,30){\usebox{\vrton}}
\put(50,40){\usebox{\vrtth}}
\put(50,50){\usebox{\vrtth}}

\put(5,5){\makebox(0,0){3}}
\put(5,15){\makebox(0,0){4}}
\put(5,25){\makebox(0,0){3}}
\put(5,35){\makebox(0,0){4}}
\put(5,45){\makebox(0,0){3}}
\put(5,55){\makebox(0,0){2}}
\put(15,5){\makebox(0,0){4}}
\put(15,15){\makebox(0,0){3}}
\put(15,25){\makebox(0,0){4}}
\put(15,35){\makebox(0,0){3}}
\put(15,45){\makebox(0,0){2}}
\put(15,55){\makebox(0,0){3}}
\put(25,5){\makebox(0,0){3}}
\put(25,15){\makebox(0,0){4}}
\put(25,25){\makebox(0,0){3}}
\put(25,35){\makebox(0,0){2}}
\put(25,45){\makebox(0,0){1}}
\put(25,55){\makebox(0,0){2}}
\put(35,5){\makebox(0,0){2}}
\put(35,15){\makebox(0,0){3}}
\put(35,25){\makebox(0,0){2}}
\put(35,35){\makebox(0,0){1}}
\put(35,45){\makebox(0,0){2}}
\put(35,55){\makebox(0,0){1}}
\put(45,5){\makebox(0,0){1}}
\put(45,15){\makebox(0,0){2}}
\put(45,25){\makebox(0,0){1}}
\put(45,35){\makebox(0,0){0}}
\put(45,45){\makebox(0,0){1}}
\put(45,55){\makebox(0,0){2}}
\put(55,5){\makebox(0,0){2}}
\put(55,15){\makebox(0,0){3}}
\put(55,25){\makebox(0,0){2}}
\put(55,35){\makebox(0,0){1}}
\put(55,45){\makebox(0,0){2}}
\put(55,55){\makebox(0,0){3}}

\sbox{\vrton}{}\sbox{\vrttw}{}\sbox{\vrtth}{}\sbox{\vrtfo}{}\sbox{\vrtfi}{}\sbox{\vrtsi}{}
\end{picture}
\setlength{\unitlength}{1 pt}
\caption{\small An arrow configuration together with the corresponding  
height function.}\label{lat}
\end{figure}

We now proceed to show how six-vertex models are related to SOS models. 
First we introduce a dual lattice. Each bond of the dual lattice now 
crosses an arrow placed on one of the bonds of the original lattice. By rotating this
arrow $ 90^{\circ}$ clockwise and placing it on the corresponding bond of the dual
lattice, we obtain an arrow configuration on the dual lattice. A height function ($ h $)
is now defined by demanding that $ h\haak{x}=h\haak{y}+1 $ if an arrow
points from $ y $ to $ x $. By fixing the height at one particular point, the
height at each point of the dual lattice is defined unambiguously. See \cite{vby}
for more details. The fact that the height difference between nearest neighbors
is always $\pm 1$ makes six-vertex models ideal models for crystal surfaces
of BCC crystals in the (100) direction.
The class of SOS models to which six-vertex models are mapped is also known as
body centered solid on solid models (BCSOS models).
In fig.\ \ref{lat} an arrow configuration on a lattice together with the corresponding height function on the
dual lattice is shown.
\section{Roughening transition in the F-model}
According to \cite{lieb2} a phase transition of Kosterlitz-Thouless type takes place in the F-model at inverse
temperature $\beta\epsilon=\ln\haak{2}$. If $\beta\epsilon>\ln\haak{2}$ the crystal
surface as described by the F-model is smooth. In this case the height-height
correlation function $G\haak{r}=\gem{\haak{h\haak{r}-h\haak{0}}^{2}}$ decays
exponentially with increasing $r$. When one takes $\beta\epsilon<\ln\haak{2}$,
the surface is in a rough phase. It can be shown that \cite{jeps}
\begin{equation}
G\haak{r}=\frac{2}{\pi\arccos\haak{1-\half\exp\haak{2\beta\epsilon}}}\ln\haak{r}
\end{equation}
The logarithmic divergence of the correlation function at large distances is caused
by thermal fluctuations in the local height of the surface with arbitrary long
wavelengths. Note that for $\epsilon>0$ the F-model has a twofold degenerate 
ground state consisting of vertex 5 on one sublattice and vertex 6 on the other 
sublattice. By introducing a staggered field this degeneracy is lifted. It has
been shown \cite{njs} that in a nonzero staggered field the F-model is in a smooth
phase for positive $\epsilon$. 
\chapter{\Rge\ for sine-Gordon type models}\label{dren}
In this chapter we will introduce the sine-Gordon type 
Hamiltonian and then show how \rge\ can be obtained for such models.
First a cut-off procedure will be introduced to define the theory.
Renormalization is carried out by first integrating over some of the degrees of freedom
of the model. The model, when formulated in terms of the remaining degrees of
freedom, will look like the original model with a lower cut-off. Finally a scale transformation
will restore the original cut-off.
\section{Effective Hamiltonians for the \stfm}\label{eff}
Since the \stfm\ can be interpreted as a solid-on-solid model (see section \ref{sos}),
it is natural to introduce a field $ h $, that describes the height of a 
surface. The Hamiltonian density of this field
must possess the same symmetries
as the \stfm. In particular we must have:
\begin{eqnarray}
F\haak{h+1,\beta s} & = & F\haak{h,-\beta s}\label{sym1}\\
F\haak{h} & = & F\haak{-h}\label{sym2}
\end{eqnarray}
Here $ s $ is the staggered field, and we have assumed that the ground state 
of the \stfm\ (for $\beta\epsilon>0 $ and $\beta s\not=0 $) corresponds to 
$ h=0 $ in the sine-Gordon model. 
From (\ref{sym1}) it follows that
\begin{equation}
F\haak{h+2,\beta s}=F\haak{h,\beta s}\label{per}
\end{equation}
(\ref{sym2}) and (\ref{per}) lead us to the  Hamiltonian density: 
\begin{equation}\label{hamd}
F\haak{h,\partial_{i}h,\partial_{ij}h,\ldots,\beta\epsilon,\beta s}
=\sum_{n=0}^{\infty}D_{n}\haak{\partial_{i}h,\partial_{ij}h,\ldots,\beta\epsilon,\beta s}
\cos\haak{n\pi h}
\end{equation}
Here $D_{n}$ is an unknown function of its arguments. According to (\ref{sym1}) 
we have
\begin{equation}\label{sym3}
D_{n}\haak{\partial_{i}h,\partial_{ij}h,\ldots,\beta\epsilon,-\beta s}=
(-1)^{n}D_{n}\haak{\partial_{i}h,\partial_{ij}h,\ldots,\beta\epsilon,\beta s}
\end{equation}
\section{The renormalization group transformation}\label{dfmd}
We will rewrite the  Hamiltonian (\ref{hamd}) as
\begin{equation}\label{sg1}
H=\sum^{\infty}_{-\infty}\int\frac{d^{2}\!x}{a^{2}}\exp(i n \pi h)
D_{n}\haak{a\partial_{i}h,a^{2}\partial_{ij}h,a^{3}\partial_{ijk}h,\ldots}
\end{equation}
We can think of the constant $ a $ as the ``lattice constant'' of the original 
microscopic  Hamiltonian. In this original model $\frac{1}{a^{2}}$ 
would be the density of degrees of freedom. The effective  Hamiltonian (\ref{sg1})  
should have the same density of degrees of freedom. The constant $ a $ appears in the 
Hamiltonian as a consequence of replacing summations by integrals and finite differences
by partial derivatives.  
We will define the Fourier transform of the field $ h(x) $ as
\begin{equation}\label{sg13}
\hat{h}\haak{k}=\frac{1}{\sqrt{V}}\int d^{2}\!x h(x)\exp\haak{-\imath k\cdot x}
\end{equation}
Here $ V $ is the volume of the system.
$ h(x) $ can then be written as
\begin{equation}\label{sg14}
h(x)=\frac{1}{\sqrt{V}}\sum_{k}\hat{h}\haak{k}\exp\haak{\imath k\cdot x}
\end{equation}
We now define a cut-off by introducing a set ($ S $) of allowed $ k $-values. 
We assume that the set $ S $ has the property:
\begin{equation}
k\in S\Rightarrow -k\in S
\end{equation}
The density of $k$-values is written as $\frac{V}{\haak{2\pi}^{2}}P\haak{k}$.
The function $ P\haak{k}$ will be called a cut-off function. We shall assume
that the cut-off is chosen such that $P\haak{0}=1$ and all derivatives of $P\haak{k}$
are zero at $k=0$.
If the volume $V$ is chosen large enough, a summation over $S$ can be replaced by an
integral:
\begin{equation}
\sum_{k\in S}F\haak{k}=V\int\frac{d^{2}\!k}{\haak{2\pi}^{2}}P\haak{k}F\haak{k}
\end{equation}
provided that the function $F$ does not correlate with the characteristic function
of $S$. In case such correlations do exist we have to replace $P\haak{k}$ by
the characteristic function of the set $S$, which we denote as $P_{c}\haak{k}$.
In general we thus have
\begin{equation}
\sum_{k\in S}F\haak{k}=V\int\frac{d^{2}\!k}{\haak{2\pi}^{2}}P_{c}\haak{k}F\haak{k}
\end{equation}
The value of $ a $ now follows by requiring $\frac{1}{a^{2}}$ to be the number
of degrees of freedom per unit volume:
\begin{equation}\label{adfi}
\frac{1}{a^{2}}=\int\frac{d^{2}\!k}{\haak{2\pi}^{2}}P\haak{k}
\end{equation}
We will denote the set of all allowed functions by $\hat{S}$. $\hat{S}$ is the set 
of all finite linear combinations of the functions $ e^{\imath k\cdot x}$
with $ k\in S $. Note that we have $\hat{h}\haak{k}=0 $ if $ h\in\hat{S}$ and $ k\not\in S $.

We now define the partition function as:
\begin{equation}\label{zdef}
Z=\int Dh e^{H}
\end{equation}
Where the measure $Dh$ on $\hat{S}$ is defined as:
\begin{equation}\label{maat}
Dh\equiv\prod_{k\in S}\frac{d\hat{h}\haak{k}}{a}R\haak{k}
\end{equation}
The function $R\haak{k}$ which occurs in the definition of the measure has to 
be chosen such that the free energy of the exactly soluble Gaussian model is
consistent with the \rgee\ for the free energy. Although the correct choice of
$R\haak{k}$ is important for a consistent description of the theory, it turns
out that its effect is equivalent to adding a constant term independent of any
couplings to the  Hamiltonian and hence doesn't influence the dependence of the
free energy on the couplings. 
\section{Renormalization}\label{dfrn}
The renormalized Hamiltonian is obtained from (\ref{sg1}) by using the Wilson-Kogut momentum shell 
integration technique \cite{ohta, kogut}. We will integrate (\ref{zdef}) over
some of the degrees of freedom, leaving us with an effective Hamiltonian ($\tilde{H}$)  
with a lower cut-off. Next a scale transformation will restore the original
cut-off and yield the renormalized Hamiltonian ($H_{R}$). 

We must now specify precisely the degrees of freedom we have to integrate over.
Since the renormalized Hamiltonian ($H_{R}$) has the same cut-off function 
$P\haak{k}$ as the original Hamiltonian ($H$), and since it is obtained from the effective 
Hamiltonian ($\tilde{H}$) after a scale transformation, $\tilde{H}$ has to have 
a cut-off function of the form $P\haak{lk}$. In terms of $l$ the scale transformation 
becomes $x\rightarrow l^{-1}x$. We thus have to construct a set $S^{\haak{1}}$
of allowed $k$-values for $\tilde{H}$, such that $S^{\haak{1}}\subset S$ and
$S^{\haak{1}}$ corresponds to the cut-off function $P\haak{lk}$. The complement
of $S^{\haak{1}}$ in $S$, denoted as $S^{\haak{2}}$, contains the degrees of freedom
we have to integrate over. We thus have to split the set $ S $ of $ k $-values
into two disjoint sets $ S^{\haak{1}}$ and $ S^{\haak{2}}$. This can be done as 
follows:
We decide to put the points $ k\in S $ and $-k\in S $ in $ S^{\haak{1}}$ 
with probability $ \frac{P\haak{lk}}{P\haak{k}}$. $ S^{\haak{2}} $ is defined as
$ S^{\haak{2}}=S-S^{\haak{1}}$.
The cut-off function for $S^{\haak{2}}$ will be denoted as $P^{\haak{2}}$, is
thus given by
\begin{equation}\label{sg7}
P^{\haak{2}}= P\haak{k}-P\haak{lk}
\end{equation}
We now construct the spaces $\hat{S}^{\haak{1}}$ and $\hat{S}^{\haak{2}}$ 
analogous to $\hat{S}$:
$\hat{S}^{\haak{i}}$ is defined as the set of all finite linear combinations
of the functions $ e^{\imath k\cdot x}$ with $ k\in S^{\haak{i}}$. We now have
\begin{equation}
\hat{S}=\hat{S}^{\haak{1}}\oplus\hat{S}^{\haak{2}}
\end{equation}
The projection of a $ h\in\hat{S}$ on $\hat{S}^{\haak{1}}$ and $\hat{S}^{\haak{2}}$ will be
denoted by $\hee $ respectively $\htw $.
The first step in the Wilson-Kogut renormalization scheme is to integrate over the field $ h^{\haak{2}} $.
After this integration one obtains an effective Hamiltonian $\tilde{H}$ which depends on $\hee$.
The final step is to restore the original cut-off by a length rescaling: 
\begin{equation}\label{sg9}
x\prime=l^{-1}x 
\end{equation} 
The renormalized field $h^{R}$ is defined as:
\begin{equation}\label{trnsh}
h^{R}\haak{x\prime}=\hee\haak{x}
\end{equation}
and the renormalized Hamiltonian $H_{R}$ is defined as:
\begin{equation}\label{trnsH}
H_{R}\haak{h^{R}}=\tilde{H}\haak{\hee}
\end{equation}
\section{Cumulant expansion} 
The integration over the field $\htw$ is performed after an 
expansion about the Gaussian model. We rewrite our Hamiltonian (\ref{sg1}) as 
\begin{equation}\label{gas}
H=H_{g}+X
\end{equation}
where $H_{g}$ is a Gaussian interaction and $X$ is a perturbation. $H_{g}$ may be split into a Gaussian interaction 
for $\hee$ and $\htw$, denoted as $H^{\haak{1}}$ respectively $H^{\haak{2}}$
\begin{eqnarray}
H_{g}=&-\frac{j}{2}\int\haak{\nabla h}^{2}d^{2}\!x=-\frac{j}{2}\sum_{k\in S}k^{2}\lhaak{h\haak{k}}^{2}&\nonumber\\
=&-\frac{j}{2}\sum_{k\in S^{\haak{1}}}k^{2}\lhaak{h\haak{k}}^{2}-\frac{j}{2}\sum_{k\in S^{\haak{2}}}k^{2}\lhaak{h\haak{k}}^{2}&\nonumber\\
=&-\frac{j}{2}\int\haak{\nabla\hee}^{2}d^{2}\!x-\frac{j}{2}\int\haak{\nabla\htw}^{2}d^{2}\!x&\nonumber\\
=&H^{\haak{1}}+H^{\haak{2}}&\label{sg10}
\end{eqnarray}
Note that for a given Hamiltonian the representation (\ref{gas}) is not 
unique because one may choose to include a Gaussian term in the perturbation 
$ X $ as well. Such a freedom of choice can sometimes be exploited in first
order calculations to improve the accuracy of calculations (see \cite{shen}).

We define the measure $Dh^{\haak{2}}$ by
\begin{equation}\label{sg15}
\int Dh^{\haak{2}}F(h)\equiv\frac{\int_{h\in\hat{S}^{\haak{2}}}DhF\haak{h}}
{\int_{h\in\hat{S}^{\haak{2}}}Dhe^{H^{\haak{2}}}}
=\frac{\int\prod_{k\in S^{\haak{2}}}d\hat{h}\haak{k}F(h)}
{\int\prod_{k\in S^{\haak{2}}}d\hat{h}\haak{k}\exp\haak{H^{\haak{2}}\haak{h^{\haak{2}}}}}
\end{equation}
where $ F(h) $ is an arbitrary function of $h$. The Gaussian average of 
a function $F$ over the field $h^{\haak{2}}$ can now be written as
\begin{equation}\label{sg16}
\gem{F\haak{h}}=\int Dh^{\haak{2}}F\haak{h}\exp
\haak{ H^{\haak{2}}\haak{h^{\haak{2}}}}
\end{equation}
We now define the effective Hamiltonian $\tilde{H}\haak{\hee}$ as follows:
\begin{equation}\label{sg17}\exp\haak{\tilde{H}\haak{\hee}}=K\int Dh^{\haak{2}}\exp\haak{H}
\end{equation}
Here $K$ is a constant. To determine $H_{R}$ one simply has to rescale $\tilde{H}$
(see (\ref{sg9}), (\ref{trnsh}) and (\ref{trnsH})).
To fix the constant K, one has to express $ H_R $ and $ H $ in the same functional form
and then require the constant terms to be equal.
From (\ref{gas}), (\ref{sg10}), (\ref{sg16}) and (\ref{sg17}) it follows
\begin{equation}\label{sg18}
\tilde{H}=\ln\haak{K}+ H^{\haak{1}}+\ln\gem{\exp\haak{X}}
\end{equation}
To second order in $X$, (\ref{sg18}) can be written as
\begin{equation}\label{cum}
\tilde{H}=\ln\haak{K}+ H^{\haak{1}}+\gem{X}+\half\gem{\haak{X-\gem{X}}^{2}}+\ldots
\end{equation}
This expansion is known as the cumulant expansion. For the general form of this 
expansion, see \cite{cumul}.
\section{Diagrammatic expansion}
It turns out that the terms in the cumulant expansion can be represented as
amplitudes of Feynman-diagrams. In these diagrams the correlation function
of the field $\htw$ plays the role of the propagator. In section \ref{dcor} we show
that it takes the form:
\begin{equation}\label{hr1}
G\haak{x}=
\frac{1}{jV}\sum_{k\in S^{\haak{2}}}\frac{\exp\haak{\imath k
\cdot x}}{k^2}=
\frac{1}{j}\int\frac{d^{2}\!k}{(2\pi)^{2}}P_{c}^{\haak{2}}\haak{k}\frac{\exp\haak{\imath k
\cdot x}}{k^2}
\end{equation}
where $ P_{c}^{\haak{2}} $ is the characteristic function of the set $S^{\haak{2}}$.
The amplitudes of Feynman-diagrams we will encounter later can be expressed
as integrals of products of propagators. We have to be carefull with replacing
$P_{c}^{\haak{2}}$ by $P^{\haak{2}}$ in such cases. E.g.\ we have
\begin{equation}\label{subt}
\int d^{2}x\ahaak{G\haak{x}}^{2}=\frac{1}{j^{2}}\int\frac{d^{2}k}{\haak{2\pi}^{2}}
\frac{P^{\haak{2}}\haak{k}}{k^{4}}
\end{equation}
It is not difficult to 
see that $G\haak{0}$ is universal:
\begin{equation}\label{hr3}
G(0)=\frac{1}{2\pi j}\ln\haak{l}
\end{equation}
We now consider the case of an infinitesimal cut-off change:
\begin{equation}\label{hr7}
l^{-1}=1-\epsilon
\end{equation}
(\ref{sg9}) becomes 
\begin{equation}\label{hr8}
x'=(1-\epsilon)x
\end{equation}
We now associate $\epsilon$ with an infinitesimal increase in a rescaling parameter $t$.
The renormalization process then generates one parameter families of Hamiltonians 
$H\haak{t}$. The \rge\ can then be written as
\begin{equation}\label{dhdt}
\frac{dH}{dt}=\mbox{coefficient of $\epsilon$ in}\: H_{R}
\end{equation}
The parameter $t$ is related to a length transformation:
\begin{equation}\label{resc}
x\haak{t}=e^{-t}x\haak{0}
\end{equation}
Instead of the Hamiltonian it is often more convenient to write the \rge\ in 
terms of the Hamiltonian density. We shall denote the effective Hamiltonian 
density corresponding to the effective Hamiltonian$\tilde{H}$ as $\tilde{F}$:
\begin{equation}\label{feff}
\tilde{H}\haak{\hee}=\int d^{2}x\tilde{F}\haak{\hee,\partial_{i}\hee}
\end{equation}
The renormalized Hamiltonian density, denoted as $\tilde{F}$, can thus be expressed 
in terms of $\tilde{F}$ by rewriting (\ref{feff}) in terms of $h^{R}$:
\begin{equation}
\begin{array}{ll}
H_{R}\haak{h^{R}}\equiv\tilde{H}\haak{\hee}&=\int d^{2}x\tilde{F}\haak{\hee,\partial_{i}\hee}\\
&=\int d^{2}x'\haak{1+2\epsilon}\tilde{F}\haak{h^{R},\haak{1-\epsilon}\partial_{i}h^{R}\cdots}
\end{array}
\end{equation}
where in the last line we used the transformation $x'=\haak{1-\epsilon}x$ and
$h^{R}\haak{x'}=\hee\haak{x}$. The renormalized Hamiltonian density ($F_{R}$)
can thus be expressed as
\begin{equation}\label{frfeff}
F_{R}\haak{h^{R}}=\haak{1+2\epsilon}\tilde{F}\haak{h^{R},\haak{1-\epsilon}\partial_{i}h^{R}\cdots}
\end{equation}
The \rge\ can thus be expressed as
\begin{equation}\label{dfdt}
\frac{dF}{dt}=\mbox{coefficient of $\epsilon$ in}\: F_{R}
\end{equation}

We now proceed with the derivation of the Feynman-rules for the cumulant expansion 
(\ref{cum}). It is convenient to derive these rules first for the term $\gem{e^{X}}$.
We shall see that $\ln\gem{e^{X}}$ is obtained by summing over connected diagrams only.
Let $ F\haak{h,\partial_{i}h,\partial_{ij}h,\ldots} $ be the non-Gaussian part of 
the Hamiltonian density in (\ref{sg1}). We can then write: 
\begin{equation}\label{form}
\frac{1}{n!}\gem{X^{n}}=\frac{1}{n!}\int\prod_{k=1}^{n}d^{2}\!x_{k}\gem{\prod_{k=1}^{n}F\haak{h\haak{x_{k}},\lhaakr{\partial_{i}h}_{x_{k}},\lhaakr{\partial_{ij}h}_{x_{k}},\ldots}}
\end{equation}
We can evaluate (\ref{form}) by writing $ F $, considered as a function of the 
field $ h $ and its derivatives, as a Fourier integral. We will define a Fourier transform
of $ F $ as follows:
\begin{equation}\label{form2}
\begin{array}{l}
\hat{F}\haak{\gamma,\gamma_{i},\gamma_{ij},\ldots}=
\int\frac{dh}{2\pi}\int\prod_{i}\frac{d\partial_{i}h}{2\pi}
\int\prod_{ij}\frac{d\partial_{ij}h}{2\pi}\ldots\\
F\haak{h,\partial_{i}h,\partial_{ij}h,\ldots}
e^{-\imath\rhaak{\gamma h+\gamma_{i}\partial_{i}h 
+\gamma_{ij}\partial_{ij}h \ldots}}\\
\end{array}
\end{equation}
The integrals in (\ref{form2}) are from $-\infty $ to $\infty $.
$ F $ can now be written as
\begin{equation}\label{fund1}
\begin{array}{l}
F\haak{h,\partial_{i}h,\partial_{ij}h,\ldots}=
\int d\gamma\int\prod_{i}d\gamma_{i}\int\prod_{ij}d\gamma_{ij}\ldots\\
\hat{F}\haak{\gamma,\gamma_{i},\gamma_{ij},\ldots} 
e^{\imath\rhaak{\gamma h+\gamma_{i}\partial_{i}h 
+\gamma_{ij}\partial_{ij}h \ldots}}\\
\end{array}
\end{equation}

The next step is to substitute the representation (\ref{fund1}) for
the Hamiltonian density in (\ref{form}). To facilitate this, it is convenient to
introduce multi-indices. The term in the exponent in (\ref{fund1})
can be written as
\begin{equation}\label{mltin1}
\gamma h+\sum_{k=1}^{\infty}\gamma_{i_{1},\ldots,i_{k}}\partial_{i_{1},\ldots,i_{k}}h
\end{equation}
A tuple of $k$ indices, as in the summation in (\ref{mltin1}), can be treated as a single
index. Such an index is called a multi-index. A tuple of $k$ indices will be
written as $\haak{k}$. We can thus rewrite (\ref{mltin1}) as
\begin{equation}\label{mltin2}
\sum_{k=0}^{\infty}\gamma_{\haak{k}}\partial_{\haak{k}}h
\end{equation}
Note that repeated multi-indices are only summed over while keeping the number
of indices contained in the multi-index constant. See section \ref{dfm} for all
the conventions on multi-indices. Inserting (\ref{fund1}) in (\ref{form}) gives
\begin{equation}\label{voor}
\frac{1}{n!}\gem{X^{n}}=\frac{1}{n!}\int\haak{\prod_{j=1}^{n}d^{2}\!x_{j}d\ahaak{\gamma^{\haak{j}}}}
\haak{\prod_{j=1}^{n}\hat{F}\haak{\ahaak{\gamma^{\haak{j}}}}}\gem{e^{\imath\sum_{j=1}^{n}\gamma^{\haak{j}}_{\haak{k}}\partial_{\haak{k}}h\haak{x_{j}}}}
\end{equation}
Note that the term in the exponent in (\ref{voor}) can be interpreted as the
action of a distribution (i.e.\ a linear functional) on the field $h$.
The action of a distribution $ T $ on 
a function $ h $ is denoted as $ T h $. See section \ref{dfd} of the appendix 
for a precise definition of distributions. The problem is thus to evaluate 
\begin{equation}
\gem{e^{\imath T h}}
\end{equation}
for a general distribution $T$.
In chapter \ref{dfund2} of the appendix it is shown that
\begin{equation}\label{fund2}
\gem{e^{\imath T h}}=
e^{-\frac{1}{2}T_{x}T_{y}G\haak{x-y}}e^{\imath T\hee}
\end{equation}
Here $ T $ denotes a distribution, $ T_{x} $ and $ T_{y} $ act as $ T $ 
on $ G\haak{x-y} $ considered as a function of $x$ respectively $y$ ($x$ and $y$ are
thus ``dummy''-variables). 

We now have to expand (\ref{fund2}) in powers of the propagator, and substitute the result
in (\ref{voor}). The distribution $T$ in (\ref{fund2}) is defined as follows:
First we define the distribution $T^{\haak{j}}$ as
\begin{equation}
T^{\haak{j}}h=\sum_{k=0}^{\infty}\gamma_{\haak{k}}^{\haak{j}}\partial_{\haak{k}}h\haak{x_{j}}
\end{equation}
The distribution $T$ is then defined as
\begin{equation}
T=\sum_{j=1}^{n}T^{\haak{j}}
\end{equation}
The $L^{\mbox{th}}$ order term in the propagator in the integrand of (\ref{voor}) becomes
\begin{equation}\label{cum1}
\frac{1}{n!}\frac{\haak{-1}^{L}}{2^{L}L!}\rhaak{T_{x}T_{y}G\haak{x-y}}^{L}
\haak{\prod_{j=1}^{n}\hat{F}\haak{\ahaak{\gamma^{\haak{j}}}}}e^{\imath T\hee}
\end{equation}
Each term in the expansion of $\rhaak{T_{x}T_{y}G\haak{x-y}}^{L}$ can be represented
diagrammatically. We first perform a trivial step:
\begin{equation}\label{prop}
\rhaak{T_{x}T_{y}G\haak{x-y}}^{L}=\prod_{p=1}^{L}T_{x}T_{y}G\haak{x-y}
\end{equation}
Each term in the expansion of the product can be represented diagrammatically as follows.
Draw the $N$ points $x_{j}$. If we choose from the $p^{\mbox{th}}$ term in the
product the term $\gamma^{\haak{r}}_{\haak{m}}\partial_{\haak{m}}^{\haak{x_{r}}}$ from
$T_{x}$ and the term $\gamma^{\haak{s}}_{\haak{n}}\partial_{\haak{n}}^{\haak{x_{s}}}$ from 
$T_{y}$, we draw an oriented line from $x_{r}$ to $x_{s}$, we label the line
with the value of $p$, and at the points $x_{r}$ and $x_{s}$ we put the labels
$\haak{m}$ respectively $\haak{n}$ on the line. We repeat this for all values of $p$.
There is now a one to one correspondence between the set of all possible terms
in the expansion of the product and the set of labelled diagrams. The amplitude of a labelled diagram
is obtained by inserting the appropriate product of the $\gamma$'s and the derivatives
of the propagators in (\ref{cum1}). We see that the integrals over the $\gamma$'s
result in a factor 
\begin{equation}\label{vertex}
\frac{1}{\imath^{r}}\lhaakr{\frac{\partial^{r}F}{\partial\haak{\partial_{\haak{m_{1}}}h}\cdots\partial\haak{\partial_{\haak{m_{r}}}h}}}_{h=\hee}
\end{equation}
for each vertex where $r$ lines, labelled by $\haak{m_{1}}\ldots\haak{m_{r}}$, come together.
The product of the factors $\frac{1}{\imath^{r}}$ at each vertex will precisely cancel the factor
$\haak{-1}^{L}$ in (\ref{cum1}), because each propagator gives rise to two factors
$\frac{1}{\imath}$ and there are $L$ propagators. We can simplify matters further by
omitting all labels, except the multi-indices at both ends of each propagator, 
in a labelled diagram. The amplitude of such a Feynman diagram is given by the sum 
of all the corresponding labelled diagrams. 
It is convenient to define a propagator $G_{\haak{n},\haak{m}}\haak{p-q}$ as
\begin{equation}\label{prop2}
G_{\haak{n},\haak{m}}\haak{p-q}\equiv\partial^{\haak{p}}_{\haak{n},x}
\partial^{\haak{q}}_{\haak{m},y}G\haak{x-y}
\end{equation}
Since all labelled diagrams corresponding to a nonvanishing Feynman diagram
make identical contributions, we simply have to multiply the amplitude of
one diagram by the number of ways of labeling a Feynman diagram, to obtain its
amplitude (relabeling the vertices will change the amplitude of a diagram,
but when integrated over all positions of the vertices, all diagrams obtained
from each other by a relabeling of the vertices will, of course, make identical
contributions). This amplitude then has to be integrated over all the $x_{j}$. 
We shall denote the number of ways of orienting the propagators,
labeling the propagators and the vertices by respectively $N_{1}$, $N_{2}$ and $N_{3}$.
Since the multi-indices have to be summed over, two labelings of the propagators 
will not be considered distinct if the only difference is a permutation of the multi-indices.
Two labelings of the vertices are considered distinct if it is not possible 
to transform one labeling into the other by a relabeling of the propagators.
We then have
\begin{equation}\label{symm1}
N_{1}=2^{L-k}
\end{equation}
where $k$ is the number of lines that have both there ends connected to the same 
vertex,
\begin{equation}\label{symm2}
N_{2}=\frac{L!}{\prod_{r}k_{r}!}
\end{equation}
where the product is over all ordered pairs of vertices, and $k_{r}$ denotes the
number of propagators connecting the pair $r$ and
\begin{equation}\label{symm3}
N_{3}=\frac{n!}{S}
\end{equation}
where $S$ is the order of the symmetry group of the Feynman diagram. Using (\ref{cum1}),
(\ref{vertex}), (\ref{prop2}), (\ref{symm1}), (\ref{symm2}) and (\ref{symm3}),
we see that the Feynman rules for $\gem{e^{X}}$ are as follows:
\begin{enumerate}
\item To compute the contribution that is $n^{\mbox{th}}$ order in $X$ and $L^{\mbox{th}}$ order in $\frac{1}{j}$,
draw all topological distinct Feynman diagrams with $n$ vertices and $L$ lines.
\item Label both ends of each line by arbitrary multi-indices.
\item For each vertex there is a term:
\begin{equation}\label{fvertex}
\lhaakr{\frac{\partial^{r}F}{\partial\haak{\partial_{\haak{m_{1}}}h}\cdots\partial\haak{\partial_{\haak{m_{r}}}h}}}_{h=\hee}
\end{equation}
where the $\haak{m_{i}}$ are the multi-indices on the lines at the vertex
and the derivative is evaluated at the coordinates of the vertex.
\item Each line labelled with the multi-indices $\haak{m}$ and $\haak{n}$ corresponds 
to the propagator $G_{\haak{n},\haak{m}}\haak{p-q}$:
\begin{equation}\label{fprop}
G_{\haak{n},\haak{m}}\haak{p-q}=\partial^{\haak{p}}_{\haak{n},x}
\partial^{\haak{q}}_{\haak{m},y}G\haak{x-y}
\end{equation}
where $p$ and $q$ are the coordinates of the vertices connected by the line.
\item For each line that has both its ends connected to the same point there is a 
factor $\half$.
\item For each pair of vertices connected by $k$ lines there is a factor $\frac{1}{k!}$.
\item There is a factor $\frac{1}{S}$, where $S$ is the order of the symmetry group
acting on the vertices of the diagram.
\item Integrate over all coordinates of the vertices, and sum over all multi-indices.
\end{enumerate}
We will now show that $\ln\gem{e^{X}}$ is precisely the sum of all connected diagrams.
We assume that all connected diagrams are enumerated in some arbitrary order.
Let $C_{i}$ be the amplitude of the $i^{\mbox{th}}$ connected diagram. Using
the above Feynman rules, we can write:
\begin{equation}
\gem{\exp\haak{X}}=\sum_{\ahaak{n_{i}}}\prod_{i=1}^{\infty}\frac{C_{i}^{n_{i}}}{n_{i}!}
=\prod_{i=1}^{\infty}\sum_{n=0}^{\infty}\frac{C_{i}^{n}}{n!}=\exp\haak{\sum_{i=1}^{\infty}C_{i}}
\end{equation}
\section{Evaluation of diagrams}
\begin{figure}
\begin{center}
\setlength{\unitlength}{0.01\textwidth}
\begin{picture}(40,20)
\put(10,0){\makebox(0,0){$\bullet$}}
\put(30,5){\circle{20}}
\put(30,0){\makebox(0,0){$\bullet$}}
\end{picture}
\caption{{\small The two Feynman diagrams corresponding to the first order
cumulant.}}\label{dia1}
\setlength{\unitlength}{1 pt}
\end{center}
\end{figure}
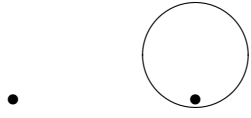

There are two diagrams, see fig.\ \ref{dia1}, contributing to the first order 
cumulant. Using the Feynman rules derived above it is a simple matter to evaluate 
the amplitude of these diagrams. 
In the case of a Hamiltonian density $F\haak{h, \partial_{i} h\cdots}$
one obtains to first order the effective Hamiltonian density $\tilde{F}$
\begin{equation}
\tilde{F}\haak{\hee,\partial_{i}\hee\cdots}=F\haak{\hee,\partial_{i}\hee\cdots}+
\half\sum_{l,m}\rhaakr{\frac{\partial^{2}F}{\partial_{\haak{k}}h\partial_{\haak{l}}h}
}_{h=\hee}G_{\haak{l},\haak{m}}\haak{0}
\end{equation}
where the sum over $l$ and $m$ is from $0$ to $\infty$.
To obtain the \rgee\ for the Hamiltonian density from this, we have to perform
a rescaling $x\rightarrow\haak{1-\epsilon}x$ (see (\ref{frfeff})) and use 
(\ref{dfdt}). These equations yield the first order \rgee\ for the Hamiltonian 
density:
\begin{equation}\label{o1}
\frac{dF}{dt}=2F-\sum_{k=1}^{\infty}k\frac{\partial F}{\partial\haak{\partial_{\haak{k}}h}}
\partial_{\haak{k}}h+\half\sum_{k,l}
\frac{\partial^{2}F}{\partial\haak{\partial_{\haak{k}}h}
\partial\haak{\partial_{\haak{k}}h}}
\frac{G_{\haak{k},\haak{l}}}{\epsilon}
\end{equation}
The sum over $k$ and $l$ is again from $0$ to $\infty$. The quantity 
$G_{\haak{k},\haak{l}}\haak{0}$ is universal (i.e.\ independent of the form of the cut-off
function $P\haak{k}$) when $k=l=0$ or $k+l=2$. 
It is not difficult to derive the result:
\begin{equation}
G_{\haak{l},\haak{m}}\haak{0}=\haak{-1}^{\frac{l-m}{2}}\frac{1}{2\pi j}
\frac{1}{2^{\frac{l+m}{2}}\haak{\frac{l+m}{2}}!}A_{l+m}C_{\haak{l},\haak{m}}
\end{equation}
Here $A_{n}$ is zero if $n$ is odd, else we have
\begin{equation}
A_{n}=\int_{0}^{\infty}d\lhaak{k}\lhaak{k}^{n-1}\haak{P\haak{\lhaak{k}}
-P\haak{\lhaak{\haak{1+\epsilon}k}}}
\end{equation}
In particular we have:
\begin{equation}
\begin{array}{lll}
A_{0}&=&\epsilon\\
A_{2}&=&\frac{4\pi}{a^{2}}\epsilon
\end{array}
\end{equation}
The tensor $C_{\haak{l},\haak{m}}$ is a contraction operator. For an arbitrary
tensor $T_{\haak{l},\haak{m}}$, $T_{\haak{l},\haak{m}}C_{\haak{l},\haak{m}}$
is the sum of all contractions of the indices contained in $\haak{l}$ and
$\haak{m}$
We can write $C_{\haak{l},\haak{m}}$ explicitly
as a sum of products of Kronecker delta's:
\begin{equation}
C_{\haak{l},\haak{m}}={\sum_{\pi}}'\delta_{\pi\haak{i_{1}},\pi\haak{i_{2}}}\cdots
\delta_{\pi\haak{i_{l+m-1}},\pi\haak{i_{l+m}}}
\end{equation}
The sum is over all nonequivalent permutations of the indices $i_{1}\cdots i_{l+m}$.
There are thus $\frac{\haak{l+m}!}{2^{\frac{l+m}{2}}\haak{\frac{l+m}{2}}!}$ terms
in the sum. 

We now proceed with the evaluation of the higher order cumulants. 
According to the Feynman rules the $n^{\mbox{th}}$ order contribution to the 
renormalized Hamiltonian is an $n$-fold integral over the volume of a product 
of propagators and functions of the field $\hee$. 
We want to replace such an 
expression by a single integral over the volume, thus obtaining a contribution 
to the Hamiltonian density. We write the amplitude $A$ of a diagram as
\begin{equation}\label{dgr}
A=\int d^{2}x_{1}\cdots d^{2}x_{n}P\haak{x_{1}\cdots x_{n}}D\haak{\hee\haak{x_{1}}\cdots\hee\haak{x_{n}}}
\end{equation}
Here $P\haak{x_{1}\cdots x_{n}}$ is the product of propagators and
$D\haak{\hee\haak{x_{1}}\cdots\hee\haak{x_{n}}}$ denotes the product of derivatives
of Hamiltonian densities. 
It is now tempting to perform $n-1$ of the $n$ integrations in (\ref{dgr}) by
Taylor-expanding the field about one of the points $x_{1}\ldots x_{n}$
(it doesn't matter
which integrations are performed because different choices are related by a partial
integration). The problem with this approach is that it assumes that the field
$\hee$ is analytic. In reality one should expect a Taylor-expansion of the field 
to converge only in a region the size of $a$ (\ref{adfi}), because $a$ is the
distance between independent degrees of freedom of the field.

A better way to proceed is to use the so-called
operator product expansion. Before we explain how this works we will first
introduce some new terminology. A local operator is a term in the Hamiltonian
that depends only on the field in one point. The Hamiltonian density evaluated
at a certain point is an example of a local operator. The first order \rge\ for 
local operators is almost identical to that of the Hamiltonian
density. If $O\haak{h\haak{x}}$ is a local operator then
\begin{equation}\label{op}
\frac{dO}{dt}=-\sum_{k=1}^{\infty}k\frac{\partial O}{\partial\haak{\partial_{\haak{k}}h}}
\partial_{\haak{k}}h+\half\sum_{k,l}
\frac{\partial^{2}O}{\partial\haak{\partial_{\haak{k}}h}
\partial\haak{\partial_{\haak{k}}h}}
\frac{G_{\haak{k},\haak{l}}}{\epsilon}
\end{equation}
The only difference with (\ref{o1}) is that the term $2O$ doesn't appear on the
r.h.s.\ of this equation.
Eigenoperators are local operators that renormalize as
\begin{equation}\label{ego}
\frac{dO}{dt}=-\lambda O
\end{equation}
$\lambda$ is called the scale dimension of the operator $O$. 
In table \ref{tbop} we have listed a few eigenoperators with their scale dimensions.
\begin{table}
\begin{center}
\begin{tabular}{|l|c|}\hline
Eigenoperator & Scale dimension\\\hline
1&0\\\hline
$\cos\haak{\pi h}$&$\frac{\pi}{4j}$\\\hline
$a^{2}\haak{\nabla h}^{2}-\frac{1}{j}$&2\\\hline
$\haak{a^{2}\haak{\nabla h}^{2}-\frac{1}{j}}\cos\haak{\pi h}$ &$\frac{\pi}{4j}+2$\\\hline
$\cos\haak{2\pi h}$&$\frac{\pi}{j}$\\\hline
\end{tabular}
\caption{{\small Some eigenoperators and their scale dimensions relative to the
Gaussian Hamiltonian} $H_{g}=-\frac{j}{2}\int d^{2}x\haak{\nabla h}^{2}$.}\label{tbop}
\end{center}
\end{table}
By solving (\ref{ego})
for all eigenoperators one obtains a complete set of operators. All eigenoperators
can be written as a multinomial of derivatives of the field $h$ multiplied by
$\cos\haak{n\pi h}$ with $n$ an integer. We shall call an operator even (odd) 
if $n$ is even (odd).
We can now
expand any local operator in this set of eigenoperators. A product of operators
localized at different points can be considered to be local if the points
lie close to each other. This product can then be expanded in eigenoperators
localized at one point. It is clear that this expansion, known as the operator 
product expansion, can be used to replace $D\haak{\hee\haak{x_{1}}\cdots\hee\haak{x_{n}}}$
in (\ref{dgr}) by a sum of operators localized at the point $x_{1}$. Suppose all
eigenoperators are enumerated by an index $n$. The scale dimension of the $n^{\mbox{th}}$
operator will be denoted as $\lambda_{n}$. We can thus put
\begin{equation}\label{dgre}
D\haak{h\haak{x_{1}}\cdots h\haak{x_{n}}}\equiv\sum_{k}c_{k}\haak{x_{1}\cdots x_{n}}
O_{k}\haak{h\haak{x_{1}}}
\end{equation}
Note that we have replaced the field $\hee$ by $h$. To apply (\ref{dgre}) to
(\ref{dgr}) a rescaling must thus be performed first. It is important to note that
(\ref{dgre}) is an identity in the sense that the Hamiltonian to which the
l.h.s.\ is added may be identified with the Hamiltonian to which the r.h.s.\ is
added. Since 
$D\haak{h\haak{x_{1}}\cdots h\haak{x_{n}}}$ is a sum of products of eigenoperators
located at the points $x_{1}\cdots x_{n}$, all we need to know are the functions
$c_{i_{1}\cdots i_{n}, k}\haak{x_{1}\cdots x_{n}}$ (operator product expansion coefficients
) in the expansion
\begin{equation}\label{ope}
O_{i_{1}}\haak{h\haak{x_{1}}}\cdots O_{i_{n}}\haak{h\haak{x_{n}}}\equiv
\sum_{k}c_{i_{1}\cdots i_{n}, k}\haak{x_{1}\cdots x_{n}}O_{k}\haak{h\haak{x_{1}}}
\end{equation}
The operator product expansion coefficients can be determined as follows. One
demands that the replacement of the product of eigenoperators according to (\ref{ope})
commutes with a renormalization. One then obtains an equation relating 
$c_{i_{1}\cdots i_{n}, k}\haak{x_{1}\cdots x_{n}}$ to 
$c_{i_{1}\cdots i_{n}, k}\haak{\frac{x_{1}}{l}\cdots\frac{x_{n}}{l}}$.
with $l$ the rescaling factor involved in the renormalization. Now, when one
takes $x_{1}=x_{2}=\cdots =x_{n}$ the operator product expansion is trivial.
The functions $c_{i_{1}\cdots i_{n}, k}\haak{x_{1}\cdots x_{n}}$ can thus be determined by taking 
the limit $l\rightarrow\infty$. Note that that since the renormalization has to
be carried out perturbatively one obtains the operator product expansion coefficients
as an expansion in the non-Gaussian couplings. It is thus very straightforward
to find the operator product expansion coefficients to zeroth order. Higher order
contributions to the operator product expansion coefficients will again involve
nontrivial integrations over Feynman-diagrams. These diagrams must again be evaluated
using the operator product expansion. E.g.\ to find the \rge\ to second order
one has to deal with expressions as in (\ref{dgr}) with $n=2$. Since the function 
$D$ is already of second order one only has to work out the operator product 
expansion to zeroth order, which is straightforward. To third order one has
to calculate the operator product expansion coefficients in (\ref{ope}) with  
$n=3$ to zeroth order and the operator product expansion coefficients with $n=2$
to first order. The latter ones involve Feynman-diagrams in which the two operators
are connected to one of the other operators in the Hamiltonian. These diagrams
can be evaluated by again using the operator product expansion (\ref{ope}), 
but now with $n=3$ and only to zeroth order. It is clear that repeated use of
the operator product expansion allows one to obtain the \rge\ to any order.

in the next chapter, we are going to apply the theory to find the phase diagram
of the \stfm.
\chapter{Applications of the \rge}\label{apren}
Once the \rge\ are known it is a simple matter to obtain the singular part
of the free energy. In this section we shall first derive the \rgee\ for the 
free energy and then proceed to show how the singular part of the free energy 
is obtained from it. 

It is convenient to rewrite the Hamiltonian as
\begin{equation}\label{hamop}
H=-\frac{j}{2}\int d^{2}x\haak{\nabla h}^{2}+\sum_{n=1}^{\infty}y_{n}\int\frac{d^{2}x}{a^{2}}
O_{n}\haak{h\haak{x}}
\end{equation}
Here the $O_{n}$ are the eigenoperators defined by equations (\ref{op}) and
(\ref{ego}). We write the \rge\ as
\begin{equation}\label{frg}
\frac{dy_{i}}{dt}=\haak{2-\lambda_{i}}y_{i}+\sum_{k=2}^{\infty}\sum_{i_{1}\cdots i_{k}}
\lambda_{i,i_{1}\cdots i_{k}}y_{i_{1}}\cdots y_{i_{k}}
\end{equation}
Note that the $\lambda_{i}$ are the scale dimensions of the operators. Also note
that the Gaussian coupling is kept constant under renormalization. This is possible
because $a^{2}\haak{\nabla h}^{2}-\frac{1}{j}$ is an eigenoperator.
To find out how the singularity in the free energy is related to the $\lambda$'s
in this equation, we must first find the relation between the free energy
of the original and the renormalized system. Note that the renormalized
Hamiltonian satisfies the relation
\begin{equation}\label{sg20}
\frac{\exp\haak{H_{R}}}{Z_{R}}=\int_{h\in\hat{S}^{\haak{2}}}\frac{\exp\haak{H}}{Z}Dh
\end{equation}
where $Z_{R}$ and $Z$ are the partition functions for the renormalized
respectively the original system. 
Combining (\ref{sg15}) with (\ref{sg17}) and (\ref{sg20}) we get:
\begin{equation}\label{sg21}
\frac{Z_{R}}{Z}=\frac{K}{Z_{g}}
\end{equation}
Here $ Z_{g}$ is the partition function of the Gaussian model:
\begin{equation}
Z_{g}=\int_{h\in\hat{S}^{\haak{2}}}\exp\haak{H^{\haak{2}}}Dh
\end{equation}
Let $U$ be the constant contribution to $H_{R}$ from
$\ln\gem{\exp\haak{X}}$. Since the total constant contribution to $H_{R}$ 
is zero, it follows from (\ref{sg18}) that $\ln\haak{K}=-U $. (\ref{sg21}) can then be written
\begin{equation}\label{sg22}
F=\haak{1-2\epsilon}F_{R}+\frac{U}{V}+\frac{1}{V}\ln\haak{Z_{g}}
\end{equation}
where $ F $ and $ F_{R} $ are the free energy densities times $-\beta$ for the original respectively 
renormalized system.
\begin{equation}\label{fr1}
\frac{dF}{dt}-2F+c+c'=0
\end{equation}
Here $c$ is the coefficient of $\epsilon $ in $\frac{U}{V}$ 
and $c'$ is the coefficient of $\epsilon $ in $\frac{1}{V}\ln\haak{Z_{g}}$.
Since $c'$ only depends on $j$, which is kept constant under renormalization,
the effect of this term is to shift the free energy by a constant amount. We are 
thus allowed to ignore this term.
\section{The case of the \stfm}
When $\beta s=0$ and $\beta\epsilon<\ln\haak{2}$, it is known that the F-model 
renormalizes to the Gaussian model \cite{kada2, knps1, knps2}. For the latter we have
\begin{equation}\label{gaus}
H=\frac{-j}{2}\int d^{2}x\haak{\nabla h}^{2}
\end{equation}
The Gaussian coupling $j$ is a known function of the temperature of the F-model:
\begin{equation}\label{iden}
j=\half\arccos\haak{1-\half e^{2\beta\epsilon}}
\end{equation}
(\ref{iden}) is valid when $\beta\epsilon<\ln\haak{2}$ and is obtained as follows: 
The long range part of the height-height correlation 
function  $R\haak{r}\equiv\gem{\haak{h\haak{r}-h\haak{0}}^{2}}$ of both models show a logarithmic behaviour, and is thus 
invariant under horizontal scaling. This means that the amplitude of the height-height correlation function is invariant
under a renormalization. 
For the F-model one finds \cite{jeps}
\begin{equation}
R\haak{r}\sim\frac{2}{\pi\arccos\haak{1-\half\exp\haak{2\beta\epsilon}}}\ln\haak{r}
\end{equation}
In case of the Gaussian model one finds (see section \ref{hhcr})
\begin{equation}
R\haak{r}\sim\frac{1}{\pi j}\ln\haak{r}
\end{equation}
Equating the amplitudes of both correlation functions 
then leads to the identification (\ref{iden}). 

We expect that when $\beta s\approx 0$, we may replace
(\ref{hamd}) by a Hamiltonian of the form (\ref{hamop}). 
Then because of (\ref{sym3}) the $y_{n}$ multiplying even (odd)
operators will be even (odd) functions of $\beta s$.
We now assume that the $y_{n}$ in (\ref{hamop}) are analytic in some neighborhood
of $\beta s=0$. 
This implies that the $y_{n}$ corresponding to odd operators are $O\haak{\beta s}$.
Of all operators, the operator $O_{1}\haak{h}=\cos\haak{\pi h}$ has the lowest scale dimension:
\begin{equation}\label{scl}
\lambda_{1}=\frac{\pi}{4j}
\end{equation}
Between $j=\frac{\pi}{8}$ and $j=\frac{\pi}{2}$ this is the only relevant operator
(i.e.\ an initially infinitesimal $y_{1}$ increases exponentially under renormalization).
Below $j=\frac{\pi}{8}$ there are no relevant operators and above $j=\frac{\pi}{2}$
the operator $\cos\haak{2\pi h}$ also becomes relevant. Because the coupling $y_{1}$
becomes proportional to $\beta s$ in the limit $\beta s\rightarrow 0$, we expect 
the \stfm\ with an infinitesimal staggered field to be in a different phase than at
zero staggered field for those values for $\beta\epsilon$ that correspond to a 
value for $j$ between $j=\frac{\pi}{8}$ and $j=\frac{\pi}{2}$. According to (\ref{iden})
this is for $\beta\epsilon$ in the interval $\half\ln\haak{2-\sqrt{2}}<\beta\epsilon<\ln\haak{2}$.
Note that at the lower boundary $\beta\epsilon$ is negative: $\half\ln\haak{2-\sqrt{2}}\approx -0.2674$
At zero staggered field the logarithmic behaviour of the height-height correlation
function indicates that the surface is in a rough phase. If the staggered field
is turned on the model no longer renormalizes to a Gaussian model. If the staggered
field is chosen small enough we expect that under a renormalization the model
will renormalize first to a model of the form
\begin{equation}\label{hamd2}
H=\int d^{2}x\rhaak{\frac{-j}{2}\haak{\nabla h}^{2}+\frac{y_{1}}{a^{2}}\cos\haak{\pi h}}
\end{equation}
with a small value for $y_{1}$, but as we renormalize further the coupling $y_{1}$ 
will increase. Since the effect of the operator $\cos\haak{\pi h}$ in the Hamiltonian
is to favour even values of $h$, we expect to be in a smooth phase. Below $\beta\epsilon
=\half\ln\haak{2-\sqrt{2}}$ we still expect that the model will renormalize to 
(\ref{hamd}) but as we renormalize further $y_{1}$ will renormalize to zero. We
are then left with a purely Gaussian Hamiltonian which describes a rough surface.

Above $\beta\epsilon=\ln\haak{2}$ the operator $\cos\haak{2\pi h}$ becomes relevant.
Since this is an even operator its coupling is nonzero at zero staggered field.
This causes the model to no longer renormalize to the Gaussian model (as a consequence
(\ref{iden}) is not valid in this region). If $\beta\epsilon>\ln\haak{2}$
the surface is thus in a smooth phase even if $\beta s=0$. To complete the
phase diagram we must find the behaviour of the model for finite values of the 
staggered field below $\beta\epsilon=\half\ln\haak{2-\sqrt{2}}$. Before we do 
that we shall first calculate the singular part of the free energy above
$\beta\epsilon=\half\ln\haak{2-\sqrt{2}}$ at $\beta s=0$.
\section{Singular part of the free energy of the \stfm}
We shall assume that the \stfm\ can be mapped to a model of the form (\ref{hamop})
such that the couplings $y_{n}$ are analytical as a function of $\beta s$ in a 
neighborhood of $\beta s=0$.
If the free energy $F$ of the model (\ref{hamop}) is written as
\begin{equation}
F=F_{s}+F_{r}
\end{equation}
with $F_{s}$ the singular part of the free energy and $F_{r}$ the regular part of
the free energy, $F_{s}$ will satisfy the homogeneous part of (\ref{fr1}) and
$F_{r}$ will be a full solution of (\ref{fr1}). Exceptions to this rule may
arise when a critical exponent associated with the singular behaviour of the free
energy is an even integer as we shall see later. Ignoring these exceptions for 
the moment, we see that
$F_{s}$ satisfies the equation:
\begin{equation}\label{sing}
F_{s}\haak{y_{1}\haak{t},y_{2}\haak{t}\ldots}=e^{2t}
F_{s}\haak{y_{1}\haak{0},y_{2}\haak{0}\ldots}
\end{equation}
In order to see how irrelevant operators modify the singular behaviour, it is
enough to keep just one irrelevant coupling. The generalization to more irrelevant
couplings is trivial.
Suppose that for $\beta s\approx 0$ the \stfm\ model is mapped to a model 
(\ref{hamop}) with $y_{1}\haak{0}$ the coupling of the relevant
operator $\cos\haak{\pi h}$ and $y_{2}\haak{0}$ the coupling of an irrelevant 
even operator. The mapping to the model (\ref{hamop}) can then be written
\begin{equation}\label{id}
\begin{array}{lcl}
y_{1}\haak{0}&=&R_{1}\beta s+O\haak{\haak{\beta s}^{3}}\\
y_{n}\haak{0}&=&R_{n}+O\haak{\haak{\beta s}^{2}}\\
\end{array}
\end{equation}
The \rge\ (\ref{frg}) can be rewritten as 
\begin{equation}\label{reny}
\frac{dy_{i}}{dt}=a_{i}y_{i}
\end{equation}
With $a_{1}=2-\frac{\pi}{4j}>0$ and $a_{n}<0$. Higher order terms in the \rge\ have been ignored.
From (\ref{sing}) and (\ref{reny}) it then follows that
\begin{equation}\label{siny}
F_{s}\haak{y_{1}\haak{0},y_{2}\haak{0}}=e^{-2t}
F_{s}\haak{y_{1}\haak{0}e^{a_{1}t},y_{2}\haak{0}e^{a_{2}t}}
\end{equation}
Now choose $t$ such that
\begin{equation}
y_{1}\haak{0}e^{a_{1}t}=c
\end{equation}
with $c$ a constant $\not = 0$.
we can then rewrite (\ref{siny}) as
\begin{equation}\label{siny2}
F_{s}\haak{y_{1}\haak{0},y_{2}\haak{0}}=
\haak{\frac{y_{1}\haak{0}}{c}}^{\frac{2}{a_{1}}}
F_{s}\haak{c,y_{2}\haak{0}\haak{
\frac{y_{1}}{c}}^{\frac{-a_{2}}{a_{1}}}}
\end{equation}
Since we expect $F_{s}$ to be analytical as a function of $y_{2}\haak{0}$
as long as $y_{1}\haak{0}\not = 0$, we can expand:
\begin{equation}\label{ex}
F_{s}\haak{c,
y_{2}\haak{0}\haak{\frac{y_{1}}{c}}^{\frac{-a_{2}}{a_{1}}}}=
A+B y_{2}\haak{0}\haak{\frac{y_{1}}{c}}^{\frac{-a_{2}}{a_{1}}}+\ldots
\end{equation}
Inserting this into (\ref{siny2}) and using (\ref{id})
yields for the leading singularity in the free energy ($F_{1}\haak{\beta s}$):
\begin{equation}\label{cex}
F_{1}\haak{\beta s}\sim\lhaak{\beta s}^{\frac{2}{a_{1}}}
\end{equation}
while the irrelevant operator contributes a singularity ($F_{2}\haak{\beta s}$) of the form
\begin{equation}
F_{2}\haak{\beta s}\sim F_{1}\haak{\beta s}\lhaak{\beta s}^{\frac{-a_{2}}{a_{1}}}
\end{equation}
Note that $a_{1}=2-\frac{\pi}{4j}$, and $j$ is given by (\ref{iden}). As $\beta\epsilon$
approaches $\half\ln\haak{2-\sqrt{2}}$ from above $a_{1}$ tends to zero, and
the singularity in the free energy becomes weaker and weaker. What happens at 
$\beta\epsilon=\half\ln\haak{2-\sqrt{2}}$ and below is the subject of section
\ref{bel}
The above argument can easily be generalised to take account of the presence 
of more irrelevant operators and higher order terms in the 
identification (\ref{id}), 
the \rge\ (\ref{reny}) and the expansion (\ref{ex}).
By applying a general result \cite{wgn} to this case, we find that the 
free energy contains singularities of the form
\begin{equation}\label{rslt}
F_{s}\haak{\beta s}\sim\lhaak{\beta s}^{n_{0}+\frac{2}{a_{1}}-\sum_{k=2}^{\infty}\frac{n_{k}a_{k}}{a_{1}}}
\end{equation}
where the $n_{i}$ are positive integers. If the exponent becomes an even integer
we have to multiply the r.h.s.\ of (\ref{rslt}) with $\ln\lhaak{\beta s}$. We can demonstrate
this in the case of the leading singularity as follows: According to (\ref{fr1})
the \rgee\ 
for the free energy is given by
\begin{equation}\label{rfee}
\frac{dF}{dt}-2F=-c\haak{y_{1}}
\end{equation}
$c$ is an even analytical function of $y_{1}$, because $\cos\haak{\pi h}$ is an odd
operator while the constant operator is even. We are now assuming
that $\frac{2}{a_{1}}=2n$ where $n$ is an integer. (\ref{reny}) gives
\begin{equation}\label{y1t}
y_{1}\haak{t}=y_{1}\haak{0}e^{a_{1}t}
\end{equation}
If $-c$ contains a term $Ky_{1}^{2n}$, (\ref{rfee}) can be rewritten as
\begin{equation}\label{free2}
\frac{dF}{dt}-2F=K\haak{y_{1}\haak{0}}^{2n}e^{2t}
\end{equation}
From this equation it follows that 
\begin{equation}
F\haak{y\haak{t}}=K\haak{y_{1}\haak{0}}^{2n}te^{2t}+F\haak{y\haak{0}}
\end{equation}
Using (\ref{id}) and (\ref{y1t}) it then follows that
\begin{equation}
F_{s}\sim\beta s^{2n}\ln\lhaak{\beta s}
\end{equation}

It is interesting to see what happens if we let $\frac{2}{a_{1}}$ approach
the value $2n$. If we put $\frac{2}{a_{1}}=2n-\epsilon$ for small $\epsilon$,
we can rewrite (\ref{free2}) as:
\begin{equation}
\frac{dF}{dt}-2F=K\haak{y_{1}\haak{0}}^{2n}e^{\haak{2+\epsilon a_{1}}t}
\end{equation}
expanding the r.h.s.\ of this equation in powers of $\epsilon$ gives
\begin{equation}
\frac{dF}{dt}-2F=K\haak{y_{1}\haak{0}}^{2n}e^{2t}\rhaak{1+\epsilon a_{1}t+\ldots}
\end{equation}
And the singularity in the free energy can be written as
\begin{equation}
F_{s}\haak{\beta s}\sim\beta s^{2n}\rhaak{\ln\lhaak{\beta s}+\frac{\epsilon}{2}\ln^{2}\lhaak{\beta s}+\ldots}
\end{equation}
\section{The case $\beta\epsilon\leq\half\ln\haak{2-\sqrt{2}}$}\label{bel}
To complete the phase diagram we must obtain the behaviour of the model at finite
values for the staggered field. This requires us to study the effects of
higher order terms in the \rge. 
To find the most important higher order terms 
we look for terms that are second order in $y_{1}$. These terms are involved in 
the generation of operators. The most important of these terms is the one 
involved in the generation of the most relevant operator. We also look for the lowest order term in the generation of $y_{1}$
arising from interactions with operators which have as low an scale dimension
as possible.
To second order in $y_{1}$ only even operators are generated and the most relevant
of these is the operator $O_{2}\haak{h}=a^{2}\haak{\nabla h}^{2}-\frac{1}{j}$. It is also this
operator which through interaction with the operator $O_{1}$ contributes to the
generation of $O_{1}$, which is also a second order effect. Since $O_{2}$ is Gaussian
we can calculate this effect simply by perturbing the Gaussian interaction $j$.
Denoting the coupling of $O_{1}$ by $y$ and the coupling of $O_{2}$ by $-\frac{j'}{2}$, we can write
\begin{equation}\label{kstd}
\begin{array}{lcl}
\frac{dj'}{dt}&=&A\haak{j+j'}y^{2}\\
\frac{dy}{dt}&=&\haak{2-\frac{\pi}{4\haak{j+j'}}}y
\end{array}
\end{equation}
Although the function $A\haak{j}$ can be calculated using the methods developed
in the previous chapter, for our purpose we can afford to leave this function
undetermined.  
At $j=\frac{\pi}{8}$, which corresponds to $\beta\epsilon\leq\half\ln\haak{2-\sqrt{2}}$, 
the operator $O_{1}$ is marginal (i.e.\ right on the boundary between relevant 
and irrelevant). To investigate the phase diagram around this point, we put 
$j=\frac{\pi}{8}$ in (\ref{kstd}) and expand in powers of $j'$. To leading order
we find
\begin{equation}\label{kost}
\begin{array}{lcl}
\frac{dj'}{dt}&=&Ay^{2}\\
\frac{dy}{dt}&=&\frac{16}{\pi}j'y
\end{array}
\end{equation}
where $A\equiv A\haak{\frac{\pi}{8}}$. Note that these \rge\ are similar to those
for the XY model (see \cite{kt1,kt2}).
To be able to construct the phase diagram,
we must know how to relate $j'$ and $y$ to the model parameters $\beta\epsilon$
and $\beta s$ of the \stfm\ in a nonzero staggered field. According to (\ref{id})
we can put
\begin{equation}\label{id1}
\begin{array}{lcl}
y\haak{0}&=&R\haak{\beta\epsilon}\beta s+O\haak{\haak{\beta s}^{3}}\\
j'\haak{0}&=&R'\haak{\beta\epsilon}+O\haak{\haak{\beta s}^{2}}\\
\end{array}
\end{equation}
where we have used the fact that $O_{2}$ is an even operator. The function 
$R'\haak{\beta\epsilon}$ can be calculated by using the fact that at zero staggered field
the model renormalizes to a Gaussian model with coupling $j$ given by (\ref{iden}).
We thus find that 
\begin{equation}
R'\haak{\beta\epsilon}=\half\arccos\haak{1-\half e^{2\beta\epsilon}}-\frac{\pi}{8}
\end{equation}
We now put $\beta\epsilon=\half\ln\haak{2-\sqrt{2}}-u$ in (\ref{id}) and expand
to leading order. We find
\begin{equation}\label{id2}
\begin{array}{lcl}
y\haak{0}&=&R\beta s\\
j'\haak{0}&=&-\haak{\sqrt{2}-1}u\\
\end{array}
\end{equation}
where $R\equiv R\haak{\beta\epsilon=\half\ln\haak{2-\sqrt{2}}}$. According to
(\ref{kost}) it follows that $K\haak{t}$, defined as
\begin{equation}\label{cnst}
K\haak{t}=y\haak{t}^{2}-\frac{16}{\pi A}j'\haak{t}^{2}
\end{equation}
is a conserved quantity under renormalization. Above $j'=0$ all flow lines, 
irrespective of the value of $K$, renormalize to infinity. Below $j'=0$ the situation 
is different. Flow lines with negative $K$ end up on the Gaussian line, while
flow lines with positive $K$ renormalize toward infinity. The flow lines with
$K=0$ thus mark the boundary between the rough phase and the smooth phase below
$j'=0$. Using (\ref{id2}) and (\ref{cnst}), we see that the lines
\begin{equation}\label{crit}
\beta s=\pm\frac{4}{R\sqrt{\pi A}}\haak{\sqrt{2}-1}u
\end{equation}
with $u\geq 0$ are the critical lines of the \stfm. Points chosen between these
lines renormalize toward the Gaussian line, points outside this region will not.

We now proceed with a derivation the singular part of the free energy. As the
critical line is approached from the smooth side, we expect singular behaviour
of the free energy (note that points on the critical lines itself renormalize to
the point $j'=0$ on the Gaussian line, there is thus no singularity when the critical
line is approached from the rough side). Since all points on the critical lines
of the \stfm\ flow toward the same point on the Gaussian line, critical behaviour
is the same all along the critical lines. We can thus content ourselves with
a calculation of the singular behaviour of the free energy at $\beta\epsilon=\half\ln\haak{2-\sqrt{2}}$
as we let $\beta s$ approach zero. In this case we are again in the area where
the identification (\ref{id2}) and the \rge\ (\ref{kost}) are valid. The initial
values are thus $j'\haak{0}=0$ and $y\haak{0}=R\beta s$. According to (\ref{cnst})
we find that $K=R^{2}\haak{\beta s}^{2}$ for the streamline that passes through 
this point. Eliminating $y\haak{t}$ in favour of $j'\haak{t}$ and using (\ref{kost})
gives us the equation
\begin{equation}
\frac{dj'}{dt}=AK+\frac{16}{\pi}{j'}^{2}
\end{equation}
This differential equation is easily integrated:
\begin{equation}\label{tpp}
t=\frac{\pi}{4\sqrt{A\pi K}}\arctan\haak{\frac{4}{\sqrt{A\pi K}}j'\haak{t}}
\end{equation}
Using (\ref{crit}) and the fact that $K=R^{2}\haak{\beta s}^{2}$, we can
rewrite (\ref{tpp}) as
\begin{equation}
t=\frac{\pi\haak{1+\sqrt{2}}\tan\haak{\theta}}{16\lhaak{\beta s}}
\arctan\haak{\frac{\haak{1+\sqrt{2}}\tan\haak{\theta}}{\lhaak{\beta s}}j'\haak{t}}
\end{equation}
where $\theta$ is the angle at which the critical line intersects the line
$\beta s=0$.
Applying (\ref{siny}) to our case yields the leading singularity in the free energy 
$F_{s}$: 
\begin{equation}
F_{s}\haak{\beta s}\sim e^{-\frac{\pi^{2}\haak{1+\sqrt{2}}\tan\haak{\theta}}{16\lhaak{\beta s}}}
\end{equation}
We can see that the singularity is of infinite order. This is characteristic of
the Kosterlitz-Thouless transition. 
Numerical studies using transfer matrix techniques have yielded
similar results on the phase diagram of the \stfm\ \cite{erk}. 

Note that all results have been obtained by using the information present in the
behaviour of the height-height correlation function of the F-model. To obtain 
more results we clearly need more information. In the next chapter we shall  
discuss a simple method that allows one to expand the free energy about
the line $\beta\epsilon=\half\ln\haak{2}$. This expansion can be used to generate
more information about the mapping of the \stfm\ to Gaussian models.
\chapter{Expansions about free fermion models}\label{ffc}
In this final chapter we will first present Baxters solution of the \stfm\ on the
free fermion line (i.e.\ the line $\beta\epsilon=\half\ln\haak{2}$). Then we 
proceed by expanding the free energy of the \stfm\
about the free fermion line.
We shall obtain an explicit expression for the free energy to first order.
By comparing the singular behaviour of this expression to that obtained from
renormalization group arguments, we are able to verify the known behaviour 
of the Gaussian coupling to first order about $\beta\epsilon=\half\ln\haak{2}$.
To simplify the computations of the higher order terms we derive a linked cluster 
method. 
\section{Definition of free fermion models}
Baxter has solved the \stfm\ at the temperature $\beta\epsilon=\half\ln(2)$
\cite{bxt}. Later it was found that this solution could be generalized to other
models if a certain condition concerning the vertex weights is met. This condition
is called the free fermion condition because for eight-vertex models satisfying this
condition the problem leads to a problem of noninteracting fermions in the 
S-matrix formulation. Let $w_{i}$ be the vertex weight for a vertex of type 
$i$ (see fig.\ \ref{df}), then the free fermion condition for six-vertex models is:
\begin{equation}\label{fermion}
w_{1}w_{2}+w_{3}w_{4}-w_{5}w_{6}=0
\end{equation}
The weights $w_{i}$ may be chosen inhomogeneous. We now proceed by presenting
a simplified version of Baxter's solution of the \stfm.
\section{Baxter's solution of the \stfm} 
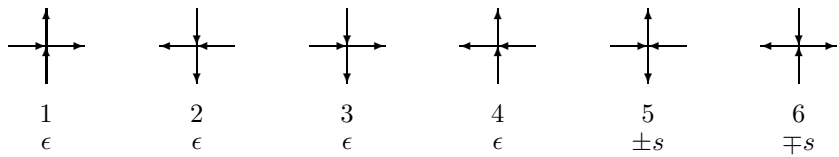
\begin{figure}
\setlength{\unitlength}{0.0165 \textwidth}
\begin{picture}(60,10)
\put(5,7.5){\vector(1,0){2.5}}
\put(2.5,7.5){\vector(1,0){2.5}}
\put(5,5){\vector(0,1){2.5}}
\put(5,7.5){\vector(0,1){2.5}}
\put(5,3){\makebox(0,0){1}}
\put(5,1){\makebox(0,0){$\epsilon $}}

\put(17.5,7.5){\vector(-1,0){2.5}}
\put(15,7.5){\vector(-1,0){2.5}}
\put(15,10){\vector(0,-1){2.5}}
\put(15,7.5){\vector(0,-1){2.5}}
\put(15,3){\makebox(0,0){2}}
\put(15,1){\makebox(0,0){$\epsilon $}}

\put(25,7.5){\vector(1,0){2.5}}
\put(22.5,7.5){\vector(1,0){2.5}}
\put(25,10){\vector(0,-1){2.5}}
\put(25,7.5){\vector(0,-1){2.5}}
\put(25,3){\makebox(0,0){3}}
\put(25,1){\makebox(0,0){$\epsilon $}}

\put(37.5,7.5){\vector(-1,0){2.5}}
\put(35,7.5){\vector(-1,0){2.5}}
\put(35,5){\vector(0,1){2.5}}
\put(35,7.5){\vector(0,1){2.5}}
\put(35,3){\makebox(0,0){4}}
\put(35,1){\makebox(0,0){$\epsilon $}}

\put(47.5,7.5){\vector(-1,0){2.5}}
\put(42.5,7.5){\vector(1,0){2.5}}
\put(45,7.5){\vector(0,1){2.5}}
\put(45,7.5){\vector(0,-1){2.5}}
\put(45,3){\makebox(0,0){5}}
\put(45,1){\makebox(0,0){$\pm s $}}

\put(55,7.5){\vector(-1,0){2.5}}
\put(55,7.5){\vector(1,0){2.5}}
\put(55,10){\vector(0,-1){2.5}}
\put(55,5){\vector(0,1){2.5}}
\put(55,3){\makebox(0,0){6}}
\put(55,1){\makebox(0,0){$\mp s $}}
\end{picture}
\setlength{\unitlength}{1 pt}
\caption{\small The six vertices and their energies. The upper and lower
signs correspond to sublattice A respectively B.}\label{df}
\end{figure}

Divide the lattice into two sublattices A and B. Choose the vertex energies as indicated
in fig.\ \ref{df}. Consider the ground state in which all A vertices are vertices of type 6,
and all B-vertices are of type 5. Any state can now be represented by drawing
lines on the lattice where the arrows point oppositely to the ground state configuration.
In terms of these lines the six vertices are represented by vertices with either
no lines, two lines at right angles, or four lines. The energies of these vertices
are respectively $-s$, $\epsilon$ and $s$. The next step is to replace the original lattice
by a decorated lattice by replacing each original vertex by a ``city'' of four internally
connected points (see fig.\ \ref{dm}).
The lines on the original lattice are regarded as dimers on the external bonds
of the decorated lattice. For any configuration on the original lattice, it is
possible to place dimers on the internal bonds of the decorated lattice, so that
the lattice becomes completely covered. Now associate to each dimer a weight as
indicated in fig.\ \ref{dm}. We now have to choose these weights such that a close-packed 
dimer problem formulated on the decorated lattice is equivalent to our original 
problem. It is a simple matter to see that for this to be the case, we can put
\begin{eqnarray}
C=D=E=F&=&e^{-\half\beta s}\\
u&=&\half\sqrt{2}e^{\half\beta s}\\
\beta\epsilon&=&\half\ln\haak{2}\label{ferm}
\end{eqnarray}
Note that (\ref{ferm}) is indeed consistent with the free fermion condition (\ref{fermion}).
\section{The Pfaffian method}
To solve the close-packed dimer problem, we use the Pfaffian method \cite{hrst, kst, mnt}.
This method is applicable whenever the lattice is planar, and works by expressing 
the partition function of the problem as the square root of the determinant of 
an antisymmetric matrix (a Pfaffian). 

A contribution to $Z^{2}$ can be written as the product of two dimer coverings
$C$ and $C'$. If $C$ connects a point $i$ with a point $j$,
we write
\begin{equation}
C\haak{i}=j
\end{equation}
It is clear that this defines a bijective map on the lattice.
$C$ and $C'$ divide the lattice into disjoint loops and pairs 
of neighboring points (bonds) as follows: If
\begin{equation}
\begin{array}{lcl}
C\haak{i_{1}}&=&i_{2}\\
C'\haak{i_{2}}&=&i_{3}\\
C\haak{i_{3}}&=&i_{4}\\
C'\haak{i_{4}}&=&i_{5}\\
C\haak{i_{5}}&=&i_{6}\\
&\vdots&\\
C\haak{i_{n-1}}&=&i_{n}\\
C'\haak{i_{n}}&=&i_{1}\\
\end{array}
\end{equation}
then the points $i_{1}\ldots i_{n}$ form a loop. If $n=2$, we don't get a loop
but instead a single bond. Note that $n$ is always even (even on
lattice types on which loops containing an odd number of points exist, the loops
generated by $C$ and $C'$ always contain an even number of points). Since
for each loop one has two choices to define the actions of $C$ and $C'$
within the loop, a given partition of the lattice in loops and bonds is 
consistent with many different configurations $C$ and $C'$. 
$Z^{2}$ can thus be calculated by summing over all partitions of the lattice 
in loops and bonds. The contribution a partition makes is given by the appropriate 
product of the weights of dimers, multiplied by a factor $2^{L}$, where $L$ is
the number of loops in the partition. If we orient each loop, and sum over all
oriented loops, the factor two for each loop can be ommitted. Now a partition
of the lattice in oriented loops and bonds defines a permutation of the lattice
points. For arbitrary points $i$ and $j$ on the lattice, we define $W_{i,j}$ as
\begin{equation}
\begin{array}{lcl}
W_{i,j}&=&0\mbox{ if $i$ and $j$ are not connected,}\\
W_{i,j}&=&\mbox{ weight of the dimer connecting } i\mbox{ and }j
\end{array}
\end{equation}
In terms of the matrix $W$, we can write:
\begin{equation}\label{zwij}
Z^{2}={\sum_{\pi}}'\prod_{j}W_{j,\pi\haak{j}}
\end{equation}
where the sum is over all permutations that contain only cycles of even lengths
(this restriction is denoted by the prime) and the product is over all lattice
points. Note that the restriction on the summation is only necessary for lattices
where loops of odd lengths exist. We now want to rewrite the r.h.s.\ of (\ref{zwij}) 
as the determinant of a matrix. This is possible if the lattice is planar, and
works as follows: One tries to factorize the missing sign of the permutation
$\pi$ ($s\haak{\pi}$) in the sum in (\ref{zwij}), so that we have
\begin{equation}\label{prm}
s\haak{\pi}=\prod_{j}s_{j,\pi\haak{j}}
\end{equation}
with the $s_{i,j}$ depending only on $i$ and $j$, and $s_{i,j}=\pm 1$ (we only need
to define the $s_{i,j}$ when $i$ and $j$ are connected). We shall see that a proper
choice of the $s_{i,j}$ allows one to lift the constraint in the summation in (
\ref{zwij}). Anticipating this result we can write: 
\begin{equation}\label{pfaff}
Z^{2}=\det R
\end{equation}
where
\begin{equation}\label{matr}
R_{i,j}=s_{i,j}W_{i,j}
\end{equation}
The $s_{i,j}$ have to be chosen such that (\ref{prm}) is valid for all permutations 
making a nonzero contribution to (\ref{zwij}). The cycles of such a permutation 
are precisely the oriented loops of even length and bonds, and they all have 
a sign of $-1$. We thus try to define the $s_{i,j}$ such that for a closed loop 
of even length or a bond consisting of the points $i_{1}\ldots i_{n}$ we have
\begin{equation}\label{frm}
\prod_{k=1}^{n}s_{i_{k},i_{k+1}}=-1
\end{equation}
where $i_{n+1}\equiv i_{1}$. The case $n=2$ yields
\begin{equation}\label{ant}
s_{i,j}=-s_{j,i}
\end{equation}
so that $R$ is antisymmetric. We can now see that permutations containing cycles 
of odd lengths make no net contribution to $\det R$ because reversing such a
cycle changes the sign of the contribution. A permutation that contains a cycle
with an odd number of points in its interior also makes no net contribution, 
because, the lattice being planar, these points are permuted amongst themselves, 
so that the permutation contains at least one cycle of odd length. We thus have
to satisfy (\ref{frm}) only for loops with an even number of points in its interior.
This is fortunate, because it is impossible to choose the $s_{i,j}$ such that (\ref{frm})
is satisfied for all loops of even lengths. The $s_{i,j}$ can, however, be chosen to satisfy
the condition:
\begin{equation}\label{frm2}
\prod_{k=1}^{n}s_{i_{k},i_{k+1}}=\haak{-1}^{r+1}
\end{equation}
where $r$ is the number of points inside the loop. It is clear that if the
$s_{i,j}$ satisfy (\ref{frm2}) for all loops we indeed have $Z^{2}=\det R$.
We now specialize to the case of the \stfm. In this case we are dealing with
the lattice shown in fig.\ \ref{dm}. Choosing the $s_{i,j}$ amounts to giving
each bond an orientation so that $s_{i,j}$ is positive if $i$ points to $j$. 
The arrows drawn on the bonds in fig.\ \ref{dm} represent such an orientation.
We will now proof that this choice of the orientations satisfies the condition
(\ref{frm2}). The proof proceeds by induction, and depends on the fact that loops 
sharing part of their boundaries may be combined to produce larger loops. Note 
that a loop can be broken down into smaller loops if and only if the loop has 
bonds in its interior. On the lattice (\ref{dm}), there are two types of loops 
that cannot be broken down. These are the loops formed by four internal bonds 
of a city, and loops connecting four cities formed by four external bonds and 
four internal bonds. For these loops it is easily verified that (\ref{frm2}) 
is true. Since any loop can be broken down into loops of the above type, we have 
to proof that if (\ref{frm2}) holds for two arbitrary loops sharing part of 
their boundaries, it also holds for the combined loop. To see this, suppose that there
are two loops ($L_{1}$ and $L_{2}$) with respectively $r_{1}$ and $r_{2}$ interior points, with a 
continuous common boundary consisting of $q$ points. 
The combined loop ($L_{3}$) will then have
$r_{3}=r_{1}+r_{2}+q-2$ interior points. The product in (\ref{frm2}) for a loop $L_{i}$
will be denoted as $s\haak{L_{i}}$. We then have
\begin{equation}\label{ind}
s\haak{L_{1}}s\haak{L_{2}}=s\haak{L_{3}}\haak{-1}^{q-1}
\end{equation}
because if we travers $L_{1}$ and $L_{2}$ in the same direction, we travers all 
the bonds of $L_{3}$, while the $q-1$ bonds on the common boundary are all traversed from both directions.
Assuming (\ref{frm2}) holds for $L_{1}$ and $L_{2}$, it follows from (\ref{ind}):
\begin{equation}
s\haak{L_{3}}=\haak{-1}^{r_{1}+r_{2}+q-1}=\haak{-1}^{r_{3}+1}
\end{equation}
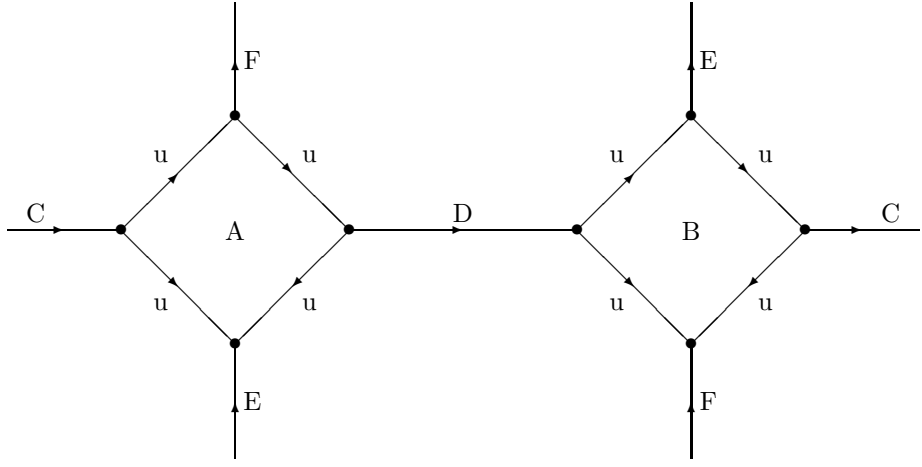
\begin{figure}
\setlength{\unitlength}{0.0625\textwidth}
\begin{picture}(16,8)
\put(1,4){\vector(1,0){0}}
\put(0,4){\line(1,0){2}}
\put(3,5){\vector(1,1){0}}
\put(2,4){\line(1,1){2}}
\put(4,7){\vector(0,1){0}}
\put(4,6){\line(0,1){2}}
\put(4,6){\line(1,-1){2}}
\put(5,5){\vector(1,-1){0}}
\put(6,4){\line(-1,-1){2}}
\put(5,3){\vector(-1,-1){0}}
\put(4,2){\line(0,-1){2}}
\put(4,1){\vector(0,1){0}}
\put(2,4){\line(1,-1){2}}
\put(3,3){\vector(1,-1){0}}
\put(0.5,4.3){\makebox(0,0){C}}
\put(2.7,5.3){\makebox(0,0){u}}
\put(5.3,5.3){\makebox(0,0){u}}
\put(5.3,2.7){\makebox(0,0){u}}
\put(2.7,2.7){\makebox(0,0){u}}
\put(4.3,1){\makebox(0,0){E}}
\put(4.3,7){\makebox(0,0){F}}
\put(2,4){\makebox(0,0){$\bullet$}}
\put(4,6){\makebox(0,0){$\bullet$}}
\put(6,4){\makebox(0,0){$\bullet$}}
\put(4,2){\makebox(0,0){$\bullet$}}
\put(11,5){\vector(1,1){0}}
\put(10,4){\line(1,1){2}}
\put(12,7){\vector(0,1){0}}
\put(12,6){\line(0,1){2}}
\put(12,6){\line(1,-1){2}}
\put(13,5){\vector(1,-1){0}}
\put(14,4){\line(-1,-1){2}}
\put(13,3){\vector(-1,-1){0}}
\put(12,2){\line(0,-1){2}}
\put(12,1){\vector(0,1){0}}
\put(10,4){\line(1,-1){2}}
\put(11,3){\vector(1,-1){0}}
\put(6,4){\line(1,0){4}}
\put(8,4){\vector(1,0){0}}
\put(14,4){\line(1,0){2}}
\put(15,4){\vector(1,0){0}}
\put(10.7,5.3){\makebox(0,0){u}}
\put(13.3,5.3){\makebox(0,0){u}}
\put(13.3,2.7){\makebox(0,0){u}}
\put(10.7,2.7){\makebox(0,0){u}}
\put(12.3,1){\makebox(0,0){F}}
\put(12.3,7){\makebox(0,0){E}}
\put(8,4.3){\makebox(0,0){D}}
\put(15.5,4.3){\makebox(0,0){C}}
\put(4,4){\makebox(0,0){A}}
\put(12,4){\makebox(0,0){B}}
\put(10,4){\makebox(0,0){$\bullet$}}
\put(12,6){\makebox(0,0){$\bullet$}}
\put(14,4){\makebox(0,0){$\bullet$}}
\put(12,2){\makebox(0,0){$\bullet$}}
\end{picture}
\setlength{\unitlength}{1 pt}
\caption{\small The ``cities'' on the decorated lattice. A and
B refer to the two sublattices. The meaning of the orientations on the bonds is
explained in the text.}\label{dm}
\end{figure}
\section{Calculation of the free energy}
\begin{figure}
\begin{center}
\setlength{\unitlength}{0.0625\textwidth}
\begin{picture}(8,8)
\put(0,4){\line(1,0){2}}
\put(2,4){\line(1,1){2}}
\put(4,6){\line(0,1){2}}
\put(4,6){\line(1,-1){2}}
\put(6,4){\line(-1,-1){2}}
\put(4,2){\line(0,-1){2}}
\put(2,4){\line(1,-1){2}}
\put(6,4){\line(1,0){2}}
\put(2,4){\makebox(0,0){$\bullet$}}
\put(4,6){\makebox(0,0){$\bullet$}}
\put(6,4){\makebox(0,0){$\bullet$}}
\put(4,2){\makebox(0,0){$\bullet$}}
\put(4,4){\makebox(0,0){$n,m$}}
\put(1.4,4.3){\makebox(0,0){$c_{n,m}$}}
\put(3.4,6.3){\makebox(0,0){$a_{n,m}$}}
\put(6.6,4.3){\makebox(0,0){$d_{n,m}$}}
\put(3.4,1.7){\makebox(0,0){$b_{n,m}$}}
\end{picture}
\end{center}
\setlength{\unitlength}{1 pt}
\caption{\small The variables {\normalsize $a_{n,m}$, $b_{n,m}$, $c_{n,m}$} and 
{\normalsize $d_{n,m}$}, associated with a city on the decorated lattice.}\label{vars}
\end{figure}
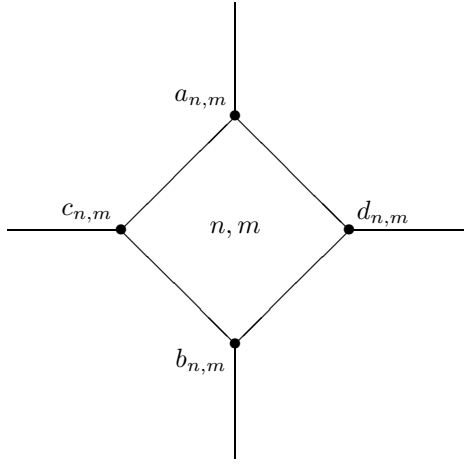

We have seen that solving the \stfm\ at $\beta\epsilon=\half\ln\haak{2}$ reduces
to the evaluation of the determinant of the matrix $R$.
To set up a perturbation theory about $\beta\epsilon=\half\ln\haak{2}$, we also
need the inverse of $R$. Both the determinant and the inverse of $R$ are easily
calculated using the following procedure: 
At each vertex $i$ on the decorated lattice, we associate a variables $x_{i}$
and $x_{i}'$.
For ${x'}_{i}$ a given set of variables, we attempt to solve the equation:
\begin{equation}\label{rcal}
R_{i,j}x_{j}={x'}_{i}
\end{equation}
We can rewrite this as follows:
Introduce coordinates $\haak{n,m}$ on the original lattice, so that increasing
$n$ ($m$) corresponds to moving to the right (upward). 
To each city on the decorated lattice we assign the coordinates of the corresponding
vertex of the original lattice. The variables $x_{i}$ and $x_{i}'$ are now given
by placing variables $a_{n,m}$, $b_{n,m}$, $c_{n,m}$, $d_{n,m}$ respectively
$a_{n,m}'$, $b_{n,m}'$,
$c_{n,m}'$, $d_{n,m}'$ for every $n$ and $m$ on the four points of the city with 
coordinates $(n,m)$ as indicated in fig.\ \ref{vars}. 
(\ref{rcal}) thus becomes
\begin{equation}\label{conf}
\begin{array}{ccccccc}
ua_{n,m}&   +&ub_{n,m}&-&Cd_{n-1,m}&=&{c'}_{n,m}\\
-ua_{n,m}&  +&ub_{n,m}&+&Cc_{n+1,m}&=&{d'}_{n,m}\\
Cb_{n,m+1}& -&uc_{n,m}&+&ud_{n,m}&=&{a'}_{n,m}\\
-Ca_{n,m-1}&-&uc_{n,m}&-&ud_{n,m}&=&{b'}_{n,m}
\end{array}
\end{equation}
We now perform a Fourier transformation on the variables:
\begin{equation}
a_{p,q}=\sum_{n,m}a_{n,m}e^{-2\pi\imath\haak{\frac{np}{N}+\frac{mq}{M}}}
\end{equation}
The Fourier transform of the other variables is defined similarly. In terms
of the Fourier transformed variables (\ref{conf}) reads:
\begin{equation}\label{ksp}
\begin{array}{ccccccc}
ua_{p,q}&+&ub_{p,q}&-&C\beta^{-p}d_{p,q}&=&{c'}_{p,q}\\
-ua_{p,q}&+&ub_{p,q}&+&C\beta^{p}c_{p,q}&=&{d'}_{p,q}\\
C\alpha^{q}b_{p,q}&-&uc_{p,q}&+&ud_{p,q}&=&{a'}_{p,q}\\
-C\alpha^{-q}a_{p,q}&-&uc_{p,q}&-&ud_{p,q}&=&{b'}_{p,q}
\end{array}
\end{equation}
Here $\alpha=e^{\frac{2\pi\imath}{M}}$ and $\beta=e^{\frac{2\pi\imath}{N}}$.
The determinant $\Delta_{p,q}$ of (\ref{ksp}) is given by:
\begin{equation}
\Delta_{p,q}=2\cosh\haak{2\beta s}+2\cos\haak{\frac{2\pi p}{N}}
\cos\haak{\frac{2\pi q}{M}}
\end{equation}
And the reduced free energy per vertex (i.e.\ the free energy times $-\beta$, 
denoted as $F$) for an infinite by infinite lattice follows:
\begin{equation}\label{frbx}
\begin{array}{l}
F=\lim_{N,M\rightarrow\infty}\frac{1}{2NM}\ln\det R=\lim_{N,M\rightarrow\infty}\frac{1}{2NM}
\sum_{p,q}\ln\Delta_{p,q}\\
=\frac{1}{8\pi^{2}}\int_{0}^{2\pi}\!\int_{0}^{2\pi}
\ln\rhaak{2\cosh\haak{2\beta s}+2\cos\haak{\theta_{1}}\cos\haak{\theta_{2}}}
d\theta_{1}d\theta_{2}\\
\end{array}
\end{equation}

\section{Singular part of the free energy}
In this section we calculate the singularity in the free energy of the
\stfm\ at $\beta s=0$ on the free fermion line. We shall use
the following method: We expand the logarithm in (\ref{frbx}), thereby
obtaining a series of the form
\begin{equation}
F\haak{\beta s}=\sum_{n=1}^{\infty}\frac{1}{\cosh^{2n}\haak{2\beta s}}
\rhaak{\frac{A_{1}}{n^{2}}+\frac{A_{2}}{n^{3}}+\cdots}
\end{equation}
Summing the series inside the
brackets term by term we obtain
\begin{equation}
F\haak{\beta s}=\sum_{k=2}^{\infty}F_{k}\haak{\beta s}
\end{equation}
with
\begin{equation}
F_{k}\haak{\beta s}=\sum_{n=1}^{\infty}\frac{1}{n^{k}\cosh^{2n}\haak{2\beta s}}
\end{equation}
As can be seen by differentiating $F_{k}$ repeatedly, the singularity in $F_{k}$ becomes
weaker as $k$ increases.

Expanding the logarithm in (\ref{frbx}) yields
\begin{equation}
\begin{array}{ccc}
F\haak{\beta s}&=&-\frac{1}{8\pi^{2}}\int_{-\pi}^{\pi}\int_{-\pi}^{\pi}d\theta_{1}d\theta_{2}
\sum_{n=1}^{\infty}\frac{\cos^{n}\haak{\theta_{1}}\cos^{n}\haak{\theta_{2}}}
{n\cosh^{n}\haak{2\beta s}}\\
&=&-\half\sum_{n=1}^{\infty}\frac{1}{2n\cosh^{2n}\haak{2\beta s}}
\rhaak{\frac{\haak{2n}!}{4^{n}n!^{2}}}^{2}
\end{array}
\end{equation}
Using
\begin{equation}
n!=n^{n}e^{-n}\sqrt{2\pi n}\exp\haak{{\sum_{k=1}^{\infty}\frac{B_{2k}}{2k\haak{2k-1}}
\frac{1}{n^{2k-1}}}}
\end{equation}
where the $B_{r}$ are the Bernoulli numbers, we find
\begin{equation}\label{fex}
F\haak{\beta s}=-\frac{1}{4\pi}\sum_{n=1}^{\infty}\frac{1}{n^{2}\cosh^{2n}\haak{2\beta s}}
\rhaak{1-\frac{1}{4n}+\frac{1}{32n^{2}}+\frac{1}{128n^{3}}-\frac{5}{2048n^{4}}+\cdots}
\end{equation}
We can find the singular part of the function $\sum_{n=1}^{\infty}\frac{1}
{n^{p}\cosh^{2n}\haak{2\beta s}}$ as follows:
Put $t=\ln\haak{\cosh^{2}\haak{2\beta s}}$. We then have to find the singular part
of the function $U_{p}\haak{t}$ with
\begin{equation}\label{udf}
U_{p}\haak{t}=\sum_{n=1}^{\infty}\frac{e^{-nt}}{n^{p}}
\end{equation}
as $t\rightarrow 0$ for $p\geq 2$. From (\ref{udf}) it follows that 
\begin{equation}\label{rec}
\frac{dU_{p+1}}{dt}=-U_{p}
\end{equation}
We denote the singular part of $U_{p}$ by $\tilde{U}_{p}$. It then follows from
(\ref{rec}) that
\begin{equation}\label{recs}
\frac{d\tilde{U}_{p+1}}{dt}=-\tilde{U}_{p}
\end{equation}
For $p=1$ the sum in (\ref{udf}) is easily evaluated:
\begin{equation}
U_{1}\haak{t}=-\ln\haak{1-e^{-t}}
\end{equation}
And we see that $\tilde{U}_{1}$ is given by
\begin{equation}\label{u1}
\tilde{U}_{1}=-\ln\haak{t}
\end{equation}
From (\ref{u1}) and (\ref{recs}) it then follows that
\begin{equation}
\tilde{U}_{p}\haak{t}=\haak{-1}^{p}\frac{t^{p-1}}{\haak{p-1}!}\ln\haak{t}
\end{equation}
Inserting this in (\ref{fex}) gives 
\begin{equation}\label{fst}
F_{s}\haak{\beta s}=-\frac{1}{4\pi}\haak{t+\frac{t^{2}}{8}+\frac{t^{3}}{192}
-\frac{t^{4}}{3072}+\cdots}\ln\haak{t}
\end{equation}
Where $F_{s}\haak{\beta s}$ is the singular part of the free energy and 
$t=2\ln\haak{\cosh\haak{2\beta s}}$. Expanding (\ref{fst}) in powers of $\beta s$ gives
\begin{equation}\label{sng}
F_{s}\haak{\beta s}=-\frac{2}{\pi}\rhaak{\haak{\beta s}^{2}-\frac{1}{6}\haak{\beta s}^{4} 
+\frac{23}{180}\haak{\beta s}^{6}-\frac{593}{5040}\haak{\beta s}^{8}+\cdots}
\ln\lhaak{\beta s}
\end{equation}
\section{Perturbation theory}\label{pbth}
We now proceed with the derivation of a perturbation theory about the free fermion
line of a 6-vertex model. The Hamiltonian of a general 6-vertex model can be defined as
follows. One assigns an energy $e\haak{p,i}$ to a vertex in state $p$ (see fig.\ \ref{df}) and position $i$. The configuration of the lattice can be specified by a
function $c$ which maps a position of a vertex to a number, $1\cdots 6$, which 
is to be interpreted as the state of the vertex at that position. The reduced
Hamiltonian ($H$) is defined to be the functional that assigns to each state
$c$ its energy times $-\beta$. We can thus write
\begin{equation}
H\haak{c}=-\beta\sum_{i}e\haak{c\haak{i},i}
\end{equation}

For $H$ a Hamiltonian of a general 6-vertex model and $H_{0}$ a 
Hamiltonian of a free fermion model a perturbation $V$ can be defined so that
we have
\begin{equation}
H=H_{0}+V
\end{equation}
The partition function $Z$ can be written as:
\begin{equation}
Z=\sum_{c}e^{H_{0}\haak{c}+V\haak{c}}=Z_{0}\gem{e^{V}}
\end{equation}
Here $Z_{0}$ is the partition function
of the free fermion model. The reduced free energy can be expressed as:
\begin{equation}\label{fermcum}
F=F_{0}+\ln\gem{e^{V}} =F_{0}+\gem{V}+\half\gem{\haak{V-\gem{V}}^{2}}+\ldots
\end{equation}
Here $F_{0}$ is the reduced free energy of the free fermion model. Now write $V=\sum_{i}V_{i}$
with $V_{i}\haak{c\haak{i}}$ a perturbation of the vertex energy times $-\beta$ at position $i$. (\ref{fermcum}) can
be rewritten as:
\begin{equation}\label{frmcm}
F=F_{0}+\sum_{i}\gem{V_{i}}+\half\sum_{ij}\rhaak{\gem{V_{i}V_{j}}-\gem{V_{i}}\gem{V_{j}}}+\ldots
\end{equation}
To compute a free fermion average $\gem{V_{i_{1}}V_{i_{2}}\ldots V_{i_{n}}}$, we
can proceed as follows: Introduce a constraint in the free fermion model by
requiring the vertices at the positions $i_{1}\ldots i_{n}$ to be in the states 
$x_{1}\ldots x_{n}$. The
partition function of this model is denoted by $Z_{i_{1}\cdots i_{n}}\haak{x_{1}\ldots x_{n}}$. 
We can then write
\begin{equation}\label{gm}
\gem{V_{i_{1}}V_{i_{2}}\ldots V_{i_{n}}}=\sum_{x_{1}\ldots x_{n}}\frac{Z_{i_{1}\cdots i_{n}}\haak{x_{1}\ldots x_{n}}
V\haak{x_{1}}\ldots V\haak{x_{n}}}{Z_{0}}
\end{equation}
It now remains to calculate $Z_{i_{1}\cdots i_{n}}\haak{x_{1}\ldots x_{n}}$.
It is convenient to reformulate this problem as follows: Denote the state of
an arrow located at the bond $j$ by $s_{j}$. Put $s_{j}=1$ if the arrow points
oppositely to the ground state configuration and $s_{j}=0$ otherwise. Define
a constrained free fermion model by requiring the arrow at the bond $j_r$ to
be in state $s_{j_{r}}$ for $1\leq r\leq m$. We then want to evaluate the partition function of 
this model, which we denote as $Z\haak{s_{j_{1}}\ldots s_ {j_{m}}}$. The idea is to
perturb the weights of the dimers on the bonds $j_{r}$ infinitesimally. We assign 
a weight $C\haak{1+\epsilon_{r}}$ to the dimer on the bond $j_{r}$. The partition 
function of the (unconstrained) free fermion model ($Z\haak{\epsilon_{1}\ldots\epsilon_{m}}$) 
can be written in terms of the constrained partition functions as:
\begin{equation}\label{part1}
\begin{array}{l}
Z\haak{\epsilon_{1}\ldots\epsilon_{m}}=\sum_{\ahaak{s}}Z\haak{s_{j_{1}}\ldots s_{j_{m}}}\prod_{k=1}^{m}\haak{1+s_{j_{k}}\epsilon_{k}}\\
=Z_{0}+\sum_{k}Z\haak{s_{j_{k}}=1}\epsilon_{k}+\sum_{k<l}Z\haak{s_{j_{k}}=1
,s_{j_{l}}=1}\epsilon_{k}\epsilon_{l}+\ldots\\
\end{array}
\end{equation}
$Z\haak{\epsilon_{1}\ldots\epsilon_{m}}$ can be calculated using (\ref{pfaff}),
by making the necessary changes to $R$. We can write:
\begin{equation}\label{mtrx}
R=R_{0}+\sum_{k=1}^{m}\epsilon_{k}R_{\haak{k}}
\end{equation}
Here $R_{0}$ is the original unperturbed matrix, $R_{\haak{k}}$ is defined as follows:
\begin{eqnarray*}
&R_{\haak{k},ij}=C\\ 
&\mbox{if $i$ and $j$ are connected by $j_{k}$ and $i$ points to $j$,}\\
&R_{\haak{k},ij}=-C\label{rpert}\\ 
&\mbox{if $i$ and $j$ are connected by $j_{k}$ and $j$ points to $i$,}\\
&R_{\haak{k},ij}=0\\ 
&\mbox{if $i$ and $j$ are not connected by $j_{k}$.}
\end{eqnarray*}
Note that the $R_{\haak{k}}$ have only two nonzero matrix elements. Inserting
(\ref{mtrx}) in (\ref{pfaff}) and expanding gives:
\begin{equation}\label{zedeps}
\begin{array}{l}
Z=\sqrt{\det R}=\sqrt{\det R_{0}}e^{\half\tr\ln\rhaak{1+\sum_{k}\epsilon_{k}R_{0}^{-1}R_{\haak{k}}}}\\
=\sqrt{\det R_{0}}\rhaakl{1+\half\sum_{k}\epsilon_{k}\tr\haak{R_{0}^{-1}R_{\haak{k}}}}\\
\rhaakr{+\frac{1}{4}\sum_{k,l}\epsilon_{k}\epsilon_{l}\rhaak{\half\tr\haak{R_{0}^{-1}R_{\haak{k}}}
\tr\haak{R_{0}^{-1}R_{\haak{l}}}-\tr\haak{R_{0}^{-1}R_{\haak{k}}
R_{0}^{-1}R_{\haak{l}}}}+\ldots}\\
\end{array}
\end{equation}
Using (\ref{zedeps}) and (\ref{part1}) we can directly read off the constrained
partition functions which have all their arguments set to $+1$ (i.e.\ all the
constrained arrows point oppositely to the ground state configuration). To calculate
a constrained partition function with some of its arguments set to $0$, we simply
have to apply the principle of inclusion and exclusion (i.e.\ a M\"{o}bius inversion
on the power set of arguments). For example consider the evaluation
of $Z\haak{s_{1},s_{2},s_{3},s_{4},s_{5}}$, with $s_{1}=s_{2}=1$ and 
$s_{3}=s_{4}=s_{5}=0$. Put $t_{3}=t_{4}=t_{5}=1$. According to the principle 
of incusion and exclusion, we can write:
\begin{eqnarray}
Z\haak{s_{1},s_{2},s_{3},s_{4},s_{5}}&=&Z\haak{s_{1},s_{2}}\nonumber\\
&&-\rhaak{Z\haak{s_{1},s_{2},t_{3}}+ Z\haak{s_{1},s_{2},t_{4}}+Z\haak{s_{1},s_{2},t_{5}}}\nonumber\\
&&+Z\haak{s_{1},s_{2},t_{3},t_{4}}+Z\haak{s_{1},s_{2},t_{3},t_{5}}+Z\haak{s_{1},s_{2},t_{4},t_{5}}\nonumber\\
&&-Z\haak{s_{1},s_{2},t_{3},t_{4},t_{5}}\label{minv}
\end{eqnarray}
\section{First order computation for the \stfm}
For the \stfm\ the expansion can be simplified. The vertex in the ground state 
at a particular point will be referred to as an a-vertex. A b-vertex is obtained
by reversing the arrows of an a-vertex. An a-vertex (b-vertex) is thus of type 
5 or 6 and has an energy of $-s$ $\haak{s}$. The constrained partition function 
corresponding to the model with one vertex constrained to be an a-vertex (b-vertex)
is denoted as $Z_{a}$ $\haak{Z_{b}}$. Note that under the transformation $s\rightarrow -s$ 
the r\^{o}le of vertices a and b are interchanged. We thus have
\begin{equation}\label{sym}
Z_{a}\haak{\beta s}=Z_{b}\haak{-\beta s}
\end{equation}
If we put $\beta\epsilon=\half\ln\haak{2}+U$ we have, according to (\ref{frmcm})
and (\ref{gm}), to first order in $U$:
\begin{equation}\label{fu}
F=F_{0}-\frac{Z-Z_{a}-Z_{b}}{Z_{0}}U+O\haak{U^{2}}
\end{equation}
Here $F$ is the reduced free energy per vertex of the \stfm. To calculate $Z_{b}$ we
only have to constrain two opposing arrows of one vertex to point oppositely
to an a-vertex. If we choose to constrain two horizontal arrows, we need
the $R^{-1}_{ij}$ for $i$ and $j$ corresponding to an $a$ or $b$ variable
on the decorated lattice. We define Green's functions $G_{x^{\haak{i}},{x^{'}}^{\haak{j}}}\haak{p,q}$ 
with $x^{\haak{i}}\in\ahaak{a,b,c,d}$ and ${x^{'}}^{\haak{j}}\in\ahaak{a',b',c',d'}$, so that the 
solution of (\ref{ksp}) can be written as
\begin{equation}
x^{\haak{i}}_{p,q}=\sum_{j}G_{x^{\haak{i}},{x^{'}}^{\haak{j}}}\haak{p,q}{x^{'}}^{\haak{j}}
\end{equation}
Solving equation (\ref{ksp}) yields
\begin{equation}\label{slksp}
\begin{array}{ccc}
G_{a,a'}\haak{p,q}&=&-\frac{1}{\Delta\haak{p,q}}\frac{e^{\half\beta s}}{2}\haak{\beta^{p}-\beta^{-p}}\\
G_{a,b'}\haak{p,q}&=&-\frac{1}{\Delta\haak{p,q}}\rhaak{\frac{e^{\half\beta s}}{2}
\haak{\beta^{p}+\beta^{-p}}+e^{-\frac{3}{2}\beta s}\alpha^{q}}\\
G_{b,a'}\haak{p,q}&=&\frac{1}{\Delta\haak{p,q}}\rhaak{\frac{e^{\half\beta s}}{2}
\haak{\beta^{p}+\beta^{-p}}+e^{-\frac{3}{2}\beta s}\alpha^{-q}}\\
G_{b,b'}\haak{p,q}&=&\frac{1}{\Delta\haak{p,q}}\frac{e^{\half\beta s}}{2}\haak{\beta^{p}-\beta^{-p}}
\end{array}
\end{equation}
An inverse Fourier transformation yields in the limit $N,M\rightarrow\infty$:
\begin{equation}\label{slconf}
\begin{array}{ccc}
G_{a,a'}\haak{n,m}&=&-\frac{e^{\half\beta s}}{8\pi^{2}}\int_{0}^{2\pi}\int_{0}^{2\pi}
d\theta_{1}d\theta_{2}\frac{e^{\imath m\theta_{1}}\haak{e^{\imath\haak{n+1}\theta_{2}}-
e^{\imath\haak{n-1}\theta_{2}}}}{\Delta\haak{\theta_{1},\theta_{2}}}\\
G_{a,b'}\haak{n,m}&=&-\frac{1}{4\pi^{2}}\int_{0}^{2\pi}\int_{0}^{2\pi}
d\theta_{1}d\theta_{2}\rhaakl{\frac{\frac{e^{\half\beta s}}{2}e^{\imath m\theta_{1}}
\haak{e^{\imath\haak{n+1}\theta_{2}}+e^{\imath\haak{n-1}\theta_{2}}}}{\Delta\haak{\theta_{1},\theta_{2}}}}\\
&&\rhaakr{+\mbox{}\frac{e^{-\frac{3}{2}\beta s}e^{\imath\haak{m+1}\theta_{1}}e^{\imath n\theta_{2}}}{\Delta\haak{\theta_{1},\theta_{2}}}}\\
G_{b,a'}\haak{n,m}&=&\frac{1}{4\pi^{2}}\int_{0}^{2\pi}\int_{0}^{2\pi}
d\theta_{1}d\theta_{2}\rhaakl{\frac{\frac{e^{\half\beta s}}{2}e^{\imath m\theta_{1}}
\haak{e^{\imath\haak{n+1}\theta_{2}}+e^{\imath\haak{n-1}\theta_{2}}}}{\Delta\haak{\theta_{1},\theta_{2}}}}\\
&&\rhaakr{+\mbox{}\frac{e^{-\frac{3}{2}\beta s}e^{\imath\haak{m-1}\theta_{1}}e^{\imath n\theta_{2}}}{\Delta\haak{\theta_{1},\theta_{2}}}}\\
G_{b,b'}\haak{n,m}&=&\frac{e^{\half\beta s}}{8\pi^{2}}\int_{0}^{2\pi}\int_{0}^{2\pi}
d\theta_{1}d\theta_{2}\frac{e^{\imath m\theta_{1}}\haak{e^{\imath\haak{n+1}\theta_{2}}-
e^{\imath\haak{n-1}\theta_{2}}}}{\Delta\haak{\theta_{1},\theta_{2}}}
\end{array}
\end{equation}
Here $\Delta\haak{\theta_{1},\theta_{2}}$ is given by
\begin{equation}
\Delta\haak{\theta_{1},\theta_{2}}=2\cosh\haak{2\beta s}+2\cos\haak{\theta_{1}}
\cos\haak{\theta_{2}}
\end{equation}
The necessary components of the matrix $R^{-1}$ can be expressed in terms 
of the Green's function $G$ as:
\begin{equation}
R^{-1}_{x_{n,m},x^{'}_{n',m'}}=G_{x,x'}\haak{n-n',m-m'}
\end{equation}\label{rgr}
Put $W_{b}=\frac{Z_{b}}{Z_{0}}$. It follows from (\ref{part1}) and (\ref{zedeps}) that
\begin{equation}\label{zb}
W_{b}=\frac{1}{4}\tr^{2}\haak{R_{0}^{-1}R_{\haak{k}}}-\frac{1}{2}\tr
\haak{R_{0}^{-1}R_{\haak{k}}R_{0}^{-1}R_{\haak{l}}}
\end{equation}
where $k$ and $l$ refer to the bonds that connect the points $a_{n,m}$ and $b_{n,m+1}$
respectively $a_{n,m+1}$ and $b_{n,m+2}$ on the decorated lattice. Using (\ref{rgr})
the first term in (\ref{zb}) $\haak{W_{b1}}$ can be written as
\begin{equation}
W_{b1}=\frac{e^{-\beta s}}{4}\rhaak{G_{b,a}\haak{0,1}-G_{a,b}\haak{0,-1}}^{2}
\end{equation}
Inserting the appropriate expressions given in (\ref{slconf}) in this equation
yields
\begin{equation}\label{zb1}
W_{b1}=\frac{1}{64\pi^{4}}\rhaak{\int_{0}^{2\pi}\int_{0}^{2\pi}d\theta_{1}d\theta_{2}
\frac{e^{-2\beta s}+\cos\haak{\theta_{1}}\cos\haak{\theta_{2}}}{\cosh\haak{2\beta s}
+\cos\haak{\theta_{1}}\cos\haak{\theta_{2}}}}^{2}
\end{equation}
The second contribution to $Z_{b}$ $\haak{W_{b2}}$ can be expressed in terms of 
the Green's function $G$ as:
\begin{equation}
\begin{array}{ccc}
W_{b2}&=&-\frac{e^{-\beta s}}{2}\rhaakl{G_{ba}\haak{0,2}G_{ba}\haak{0,0}+
G_{ab}\haak{0,0}G_{ab}\haak{0,-2}}\\
&&\rhaakr{\mbox{}-G_{aa}\haak{0,1}G_{bb}\haak{0,-1}
-G_{aa}\haak{0,-1}G_{bb}\haak{0,1}}
\end{array}
\end{equation}
Using (\ref{slconf}) it is not difficult to see that each of the four terms in
the brackets of this equation vanishes. We thus have
\begin{equation}\label{wb}
W_{b}=W_{b1}=
\frac{1}{64\pi^{4}}\rhaak{\int_{0}^{2\pi}\int_{0}^{2\pi}d\theta_{1}d\theta_{2}
\frac{e^{-2\beta s}+\cos\haak{\theta_{1}}\cos\haak{\theta_{2}}}{\cosh\haak{2\beta s}
+\cos\haak{\theta_{1}}\cos\haak{\theta_{2}}}}^{2}
\end{equation}
From (\ref{sym}) it follows that
\begin{equation}\label{wa}
W_{a}\haak{\beta s}=W_{b}\haak{-\beta s}
\end{equation}
The expression obtained by substituting (\ref{wb}) and (\ref{wa}) in (\ref{fu})
can be simplified by exploiting the fact that we only need the even part of the
function $W_{b}\haak{\beta s}$. We will denote even (odd) parts of functions
with a subscript $+$ ($-$). (\ref{fu}) can be rewritten as
\begin{equation}\label{fwbp}
F=F_{0}+\haak{2W_{b+}-1}U+\ldots
\end{equation}
(\ref{wb}) can be rewritten as
\begin{equation}\label{wi}
W_{b}=I^{2}
\end{equation}
with
\begin{equation}\label{it}
I=\frac{1}{8\pi^{2}}\int_{0}^{2\pi}\int_{0}^{2\pi}d\theta_{1}d\theta_{2}
\frac{e^{-2\beta s}+\cos\haak{\theta_{1}}\cos\haak{\theta_{2}}}{\cosh\haak{2\beta s}
+\cos\haak{\theta_{1}}\cos\haak{\theta_{2}}}
\end{equation}
From (\ref{wi}) it follows that 
\begin{equation}
W_{b+}=I_{+}^{2}+I_{-}^{2}
\end{equation}
From (\ref{it}) it easily follows that
\begin{equation}
I_{+}=\half
\end{equation}
and
\begin{equation}
I_{-}=-\half\frac{\partial F_{0}}{\partial\beta s}
\end{equation}
(\ref{fwbp}) can thus be written as
\begin{equation}\label{ou}
F=F_{0}+\half\rhaak{\haak{\frac{\partial F_{0}}{\partial\beta s}}^{2}-1}U+\ldots
\end{equation}
\section{Singular behavior in the vicinity of the free fermion line}
Equation (\ref{ou}) allows us to check the dependence of the Gaussian coupling
$j$ on $\beta\epsilon$ as given by (\ref{iden}) for $\beta\epsilon\approx\half\ln\haak{2}$.
We have seen that renormalization group arguments lead to the following leading
singularity in the reduced free energy:
\begin{equation}\label{sn1}
F_{s}\sim\haak{\beta s}^{\frac{2}{2-\frac{\pi}{4j}}}
\end{equation}
with 
\begin{equation}\label{sn2}
j=\half\arccos\haak{1-\half e^{2\beta\epsilon}}
\end{equation}
If we put 
\begin{equation}
\beta\epsilon=\half\ln\haak{2}+U
\end{equation}
in these equations, we find that
\begin{equation}
F_{s}=A\haak{U}\haak{\beta s}^{2}\rhaak{-\frac{8}{\pi}\haak{U+O\haak{U^{2}}}\ln\haak{\beta s}
+\frac{32}{\pi^{2}}\haak{U^{2}+O\haak{U^{3}}}\ln^{2}\lhaak{\beta s}+\ldots}
\end{equation}
If we compare this with (\ref{sng}), we find that the amplitude $A\haak{U}$ is
given by
\begin{equation}
A\haak{U}=\frac{1}{4U}
\end{equation}
It then follows that the amplitude of the term $\haak{\beta s}^{2}\ln^{2}\lhaak{\beta s}$ is
$\frac{8}{\pi^{2}}\haak{U+O\haak{U^{2}}}$. It is now a simple matter to verify
this using (\ref{ou}) and (\ref{sng}). From (\ref{sng}) and (\ref{ou}) it follows
that the order $U$ contribution to the singular part of the reduced free energy $F_{1}\haak{\beta s}$
can be written as
\begin{equation}
F_{1,s}\haak{\beta s}=\rhaak{B_{1}\haak{\beta s}\ln\lhaak{\beta s}+B_{2}\haak{\beta s}\ln^{2}\lhaak{\beta s}}U
\end{equation}
with $B_{1}$ and $B_{2}$ regular functions of $\beta s$. Inserting (\ref{sng})
in (\ref{ou}) gives 
\begin{equation}
B_{2}\haak{\beta s}=\frac{8}{\pi^{2}}\rhaak{\haak{\beta s}^{2}-\frac{2}{3}\haak{\beta s}^{4}
+\frac{79}{90}\haak{\beta s}^{6}+\ldots}
\end{equation}
We have thus verified (\ref{sn2}) to first order in $U$.
\section{Linked cluster expansion}\label{lce}
In this section we will derive a diagrammatic (linked cluster) method to compute 
terms in the perturbative expansion of the free energy. 
Instead of perturbing the energies of vertices 1...4, we will perturb the energy
of an a-vertex. We will consider the model with the following vertex energies:
vertices 1...4 are assigned an energy of $\epsilon_{0}$, with $\beta\epsilon_{0}=
\half\ln\haak{2}$, a-vertices are assigned an energy of $-s_{0}+v$, and b-vertices
are assigned an energy of $s_{0}$. The reduced free energy of this model is denoted as
$\tilde{F}\haak{-\beta v}$. It is not difficult to see that the reduced free energy of
the \stfm\ can be obtained from $\tilde{F}$:
\begin{equation}
F\haak{\beta\epsilon=\beta\epsilon_{0}+\half U,\beta s=\beta s_{0}+\half U}
=\tilde{F}\haak{U}-\half U
\end{equation}
We now define a Hamiltonian ($H\haak{\ahaak{J}}$) as
\begin{equation}
H\haak{\ahaak{J}}=H_{0}+\sum_{i}J_{i}V_{i}
\end{equation}
$H_{0}$ is a free fermion model and the $V_{i}$ are defined as follows. If $c$ is
an arrow configuration then $V_{i}\haak{c}$ is the increase in energy times
$-\beta$ of the vertex at position $i$ ($c_{i}$). In this case we thus have $V_{i}\haak{c}=U$
if $c_{i}$ is an a-vertex and $V_{i}\haak{c}=0$ otherwise. If we put all the
$J_{i}=1$, $H\haak{\ahaak{J}}$ becomes the Hamiltonian of the model defined above.
The partition function for $H\haak{\ahaak{J}}$ is denoted as $Z\haak{\ahaak{J}}$.
In the following we shall use the notation $J=p$ to indicate that $J_{i}=p$ for
all $i$.
An expansion about $J=0$ yields
\begin{equation}\label{zex}
Z\haak{J=1}=
\sum_{k=0}^{\infty}\frac{1}{k!}\sum_{i_{1}\cdots i_{k}}
\lhaakr{\frac{\partial^{k}Z\haak{\ahaak{J}}}
{\partial J_{i_{1}}\cdots\partial J_{i_{k}}}}_{J=0}
\end{equation}
From the definition of $Z\haak{\ahaak{J}}$ it follows that
the free fermion correlations $g\haak{i_{1}\cdots i_{k}}$, defined as
\begin{equation}
g\haak{i_{1},i_{2}\cdots i_{k}}\equiv\gem{V_{i_{1}}V_{i_{2}}\cdots V_{i_{n}}}
\end{equation}
can be expressed as
\begin{equation}
g\haak{i_{1},i_{2}\cdots i_{k}}=\frac{1}{Z\haak{J=0}}
\lhaakr{\frac{\partial^{k}Z\haak{\ahaak{J}}}
{\partial J_{i_{1}}\cdots\partial J_{i_{k}}}}_{J=0}
\end{equation}
The terms in the expansion (\ref{zex}) are thus readily expressed in terms 
of the free fermion correlations.
We have
\begin{equation}\label{zexg}
Z\haak{J=1}=Z\haak{J=0}\sum_{k=0}^{\infty}\frac{1}{k!}\sum_{i_{1}\cdots i_{k}}
g\haak{i_{1}\cdots i_{k}}
\end{equation}
To compute the free fermion correlations we define for $i_{1}\cdots
i_{k}$ $k$ points on the lattice, a constrained
free fermion model by demanding that the vertices at positions $i_{1}\cdots i_{k}$
be a-vertices. The partition function of this model is denoted as $Z_{i_{1}\cdots i_{k}}$.
In terms of $Z_{i_{1}\cdots i_{k}}$, $g\haak{i_{1}\cdots i_{k}}$ can be written
as
\begin{equation}\label{gzi}
g\haak{i_{1}\cdots i_{k}}=\frac{Z_{i_{1}\cdots i_{k}}}{Z\haak{J=0}}U^{k}
\end{equation}
From the correspondence of arrow configurations to dimer configurations, it follows
that $Z_{i_{1}\cdots i_{k}}$ can be written as the sum of all dimer configurations
such that at the cities $i_{1}\cdots i_{k}$ dimers are placed on two internal
bonds. For each of these cities there are two choices for their internal dimer 
configurations. The sum of all dimer configurations consistent with one particular
choice at each city is denoted as $Z'_{i_{1}\cdots i_{k}}$. Since $Z'_{i_{1}\cdots i_{k}}$
does not depend on the particular choice made, it is clear that we have 
\begin{equation}\label{zpdef}
Z_{i_{1}\cdots i_{k}}=2^{r}Z'_{i_{1}\cdots i_{k}}
\end{equation}
where $r$ is the number of different indices in $i_{1}\cdots i_{k}$.
For definiteness we assume that at each city $i_{p}$, for $p=1\cdots k$ dimers 
are placed on respectively
the bond connecting the points labeled by the variables $a_{i_{p}}$ and $c_{i_{p}}$,
and $d_{i_{p}}$ and $b_{i_{p}}$ (see fig.\ \ref{vars}). These bonds will be
referred to as modified bonds, and the position of these 
two bonds will be denoted as, respectively, $i_{p_{1}}$ and $i_{p_{2}}$. Dimers
on modified bonds will be referred to as modified dimers, and the modified lattice
is defined to be the decorated lattice, but with only the modified bonds. 
To compute $Z_{i_{1}\cdots i_{k}}$,
we define an inhomogeneous free fermion model by changing the weights of the 
modified bonds. At all positions $j$ we associate the variables $\epsilon_{j_{1}}$
and $\epsilon_{j_{2}}$ to, respectively, the modified bonds at $j_{1}$ and $j_{2}$.  
We now choose the weight of a dimer on a modified bond at a position $k$ 
to be a function of $\epsilon_{k}$, so that we have
\begin{equation}\label{zpex}
Z_{i_{1}\cdots i_{k}}=\mbox{coeff. of }\epsilon_{i_{1_{1}}}\epsilon_{i_{1_{2}}}
\cdots\epsilon_{i_{k_{1}}}\epsilon_{i_{k_{2}}} 
\mbox{in }Z\haak{\ahaak{\epsilon}}
\end{equation}
Here $Z\haak{\ahaak{\epsilon}}$ is the partition function of the inhomogeneous
free fermion model. We denote the weight of the dimer on a modified bond at a 
position $k$ as $w\haak{\epsilon_{k}}$. The choice $w\haak{\epsilon_{k}}=
u\haak{1+\sqrt{2}\epsilon_{k}}$ ($u$ is the original weight of the dimer) seems to 
work, but there is a problem if some of the indices in (\ref{zpex}) coincide.
To see this, consider the r.h.s.\ of (\ref{zpex}), when all the indices in this
equation are chosen differently. It is clear that a dimer configuration contributing 
to $Z\haak{\ahaak{\epsilon}}$ has a weight proportional to $\epsilon_{i_{1_{1}}}
\epsilon_{i_{1_{2}}}\cdots\epsilon_{i_{k_{1}}}\epsilon_{i_{k_{2}}}$ only if 
it consists of all the modified dimers at the positions $i_{1_{1}},i_{1_{2}\cdots}
i_{k_{1}},i_{k_{2}}$. The r.h.s.\ of (\ref{zpex}) is thus equal to $2^{k}Z'_{i_{1}
\cdots i_{k}}$, and this equals according to (\ref{zpdef}) the r.h.s.\ of 
(\ref{zpex}). It is also clear that with the present choice of the dimer weights
$Z\haak{\ahaak{\epsilon}}$ is a multilinear function of the $\epsilon$'s, so
(\ref{zpex}) is certainly not true when not all of the indices are different.
Although this doesn't seem to be a big problem, because there is no point in constraining
a certain vertex more than once, we have to sum over all indices to calculate
$Z\haak{J=1}$ and it is convenient to do so unrestrictively. The solution to this
problem is obvious. We define $w\haak{\epsilon_{k}}$ as
\begin{equation}\label{dfw}
w\haak{\epsilon_{k}}=u\haak{1+\frac{\sqrt{2}\epsilon_{k}}{1-\epsilon_{k}}}
\end{equation}

Since $Z\haak{\ahaak{\epsilon}}$ is a free fermion model, we can use (\ref{pfaff}):
\begin{equation}\label{pfaff2}
Z\haak{\ahaak{\epsilon}}=\sqrt{\det R\haak{\ahaak{\epsilon}}}
\end{equation}
with 
\begin{equation}
R\haak{\ahaak{\epsilon}}=R\haak{\epsilon=0}+\sum_{k}R_{\haak{k}}
\end{equation}          
where the sum is over all modified bonds, $\epsilon=0$ means $\epsilon_{i}=0$
for all $i$ and 
$R_{\haak{k}}$ is defined as follows. If the bond $k$ connects the points
$i$ to $j$ we put
\begin{equation}
R_{\haak{k}ij}=\pm\sqrt{2}\frac{u\epsilon_{k}}{1-\epsilon_{k}}
\end{equation}
The plus sign is chosen if the orientation on the bond is such that $i$ points 
to $j$, else the minus sign is chosen. All other matrix elements of $R_{\haak{k}}$ 
are zero. (\ref{pfaff2}) can be rewritten as
\begin{equation}\label{pert}
Z\haak{\ahaak{\epsilon}}=Z_{0}
e^{\half\tr\ln\haak{1+GB}}
\end{equation}
where $G=R^{-1}\haak{\epsilon=0}$, $Z_{0}=Z\haak{\ahaak{\epsilon=0}}$
and $B=\sum_{k}R_{\haak{k}}$. Note that $B_{ij}$ is only nonzero if $i$ and $j$
are connected by a modified bond, in which case
\begin{equation}
B_{ij}=R_{\haak{k}ij}
\end{equation}
if the modified bond $k$ connects $i$ to $j$.
Expanding the logarithm in this equation allows us to rewrite the exponent of 
the above equation as
\begin{equation}\label{exp}
\half\tr\ln\haak{1+GB}=
\half\sum_{n=1}^{\infty}\frac{\haak{-1}^{n+1}}{n}\sum_{i_{1},j_{1}\cdots i_{n},j_{n}}
\prod_{r=1}^{n}G_{i_{r},j_{r+1}}B_{j_{r+1},i_{r+1}}
\end{equation}
Here $i_{n+1}=i_{1}$ and $j_{n+1}=j_{1}$. 
Note that for a contribution to the summation on the r.h.s.\ of this equation
to be nonzero, $i_{r+1}$ must be chosen so that $j_{r+1}$ and $i_{r+1}$ are connected
by a modified bond. The summation can then be interpreted as a summation of amplitudes
of all closed paths of $n$ steps, where a single step consists of going from 
a position $i_{r}$ to an arbitrary position $j_{r+1}$, and then from $j_{r+1}$
to that point $\haak{i_{r+1}}$ which is connected to $j_{r+1}$ by a modified bond.
If $r=n$, $j_{r+1}$ is, of course, not arbitrary because of the conditions $i_{n+1}=i_{1}$ 
and $j_{n+1}=j_{1}$. Hence it follows that we are dealing with closed paths. To 
each step we associate
a weight which in the example given above equals $G_{i_{r},j{r+1}}B_{j_{r+1},i_{r+1}}$.
The amplitude of a path is the product of the weights of the steps forming the 
path. 
Each closed path can be defined as an oriented loop of which one point
plays the r\^{o}le of starting and end point. 
It is clear that the amplitude of
the path does not depend on the choice of the starting point. It also does not
depend on the orientation, because both $G$ and $B$ are antisymmetric. Instead
of summing over closed paths, we may thus sum over loops. To do this, we need
to know the number of closed paths that correspond to one loop. It is clear that
this is given by two times the number of ways one can attach a starting point 
to the loop. If the loop consists of $n$ steps, and if the winding number of the
loop is $p$ (i.e.\ traversing the loop once, means visiting the same points in 
the same order $p$ times) then there are $\frac{n}{p}$ ways of attaching an
starting point to the loop. Suppose all loops are enumerated in some arbitrary
order. We define the amplitude of the $r^{\mbox{th}}$ loop, denoted as $L_{r}$, 
as the amplitude of one of the closed paths to which it corresponds divided by
the winding number of the loop and multiplied by a sign ($\pm 1$). This sign is
the product of the following factors. For each modified bond there is a factor
$-1$, and there is an overall factor of $-1$. We can now rewrite (\ref{exp}) as
\begin{equation}\label{exp2}
\half\tr\ln\haak{1+GB}=
\sum_{r=1}^{\infty}L_{r}
\end{equation}
Inserting this in (\ref{pert})
gives
\begin{equation}\label{zexp}
\frac{Z\haak{\ahaak{\epsilon}}}{Z_{0}}=\prod_{r=1}^{\infty}e^{L_{r}}
\end{equation}
It thus follows that $\frac{Z\haak{\ahaak{\epsilon}}}{Z_{0}}$ is a sum of amplitudes
of diagrams, where a diagram is a set of loops and the amplitude of a diagram
is the product of amplitudes of the loops of which it consists multiplied by a
factor $\prod_{r=1}^{\infty}\frac{1}{k_{r}!}$ if loop $r$ occurs $k_{r}$ times in
the diagram.
Note that with our original definition of the weight function $w\haak{\epsilon}=
u\haak{1+\sqrt{2}\epsilon}$, $Z\haak{\ahaak{\epsilon}}$ would be a multilinear
function of the $\epsilon$'s. This implies that products of loops sharing bonds will not make
a net contribution (this does not, of course, depend on the choice of the 
weight function). To see an example of this consider two different loops
with amplitudes $L_{1}$ and $L_{2}$ sharing a bond. In that case there is a contribution
$L_{1}L_{2}$ to $Z\haak{\ahaak{\epsilon}}$. There is also a single loop, 
obtained by merging the two loops. This loop has an amplitude of $-L_{1}L_{2}$,
and we see that the two contributions have indeed cancelled. Similarly, contributions
from loops with winding numbers greater than one cancel against products of loops
with smaller winding numbers. Although, as we shall see later, such cancellations 
will also occur
in the expansion of $Z\haak{J=1}$, $\ln\haak{Z\haak{J=1}}$ will contain only
connected diagrams (the precise definition of ``connected'' will be given later
). Diagrams in which a bond appears more than once can make
a net contribution to $\ln\haak{Z\haak{J=1}}$ because the disconnected diagrams
against which such a contribution would cancel don't appear in the expansion of
$\ln\haak{Z\haak{J=1}}$.
To take into account the cancellation of loops with a winding number greater than
one against products of loops with lower winding numbers, we can rewrite (\ref{zexp})
as
\begin{equation}\label{zexpa}
\frac{Z\haak{\ahaak{\epsilon}}}{Z_{0}}=\prod_{r=1}^{\infty}\rhaak{1+L_{r}}
\end{equation}
Here the product is only over loops with a winding number of one. To derive this
equation directly from (\ref{exp2}) one proceeds as follows. The amplitude
of a loop is calculated as above, but without the overall minus sign. We may
then write the amplitude of a loop with winding number $p$ as
\begin{equation}
A_{r,p}=\frac{A_{r}^{p}}{p}
\end{equation}
Here $A_{r,p}$ is the amplitude of the loop with winding number $p$ and $A_{r}$
is the ``same'' loop with winding number one. We can then rewrite (\ref{exp2})
as
\begin{equation}
\half\tr\ln\haak{1+GB}=
-\sum_{r=1}^{\infty}\sum_{p=1}^{\infty}\frac{A_{r}^{p}}{p}=\sum_{r=1}^{\infty}
\ln\haak{1-A_{r}}
\end{equation}
Here the sum over $r$ is only over loops with a winding number of one.Inserting 
this in (\ref{pert}) then gives (\ref{zexpa}) (In that equation $L_{r}$ is defined
as $-A_{r}$).
With our definition
of the weight function (\ref{dfw}) a diagram contributing to $Z\haak{\ahaak{\epsilon}}$
is a multinomial in the $\epsilon$'s. To have that each diagram is a monomial,
we split each modified bond on the decorated lattice into an infinite number
of bonds, each of them connecting the same two points as the original bond.
These bonds are all assigned the same orientation as the original bond. The weights
of these bonds are chosen by assigning to each bond a (different) term 
in the expansion of (\ref{dfw}) in powers of $\epsilon$. It now follows that
(\ref{zexp}) still holds on this new lattice. We can see this as follows. Let 
all loops on the new lattice be enumerated. The amplitude of the $k^{\mbox{th}}$ 
loop will be denoted as $\tilde{L}_{k}$. Since the sum of the
weights of the new bonds replacing one modified bond equals the weight of the
modified bond, we can write
\begin{equation}
\sum_{r=1}^{\infty}L_{r}=\sum_{k=1}^{\infty}\tilde{L}_{k}
\end{equation}
(\ref{zexp}) can thus be rewritten as
\begin{equation}\label{zexpn}
\frac{Z\haak{\ahaak{\epsilon}}}{Z_{0}}=\prod_{r=1}^{\infty}e^{L_{r}}=
e^{\sum_{r=1}^{\infty}L_{r}}=e^{\sum_{k=1}^{\infty}\tilde{L}_{k}}=
\prod_{k=1}^{\infty}e^{\tilde{L}_{r}}
\end{equation}
Note that the cancellation
of diagrams discussed above implies that a diagram on the new lattice which 
uses more than bond connecting the same points will make no net contribution.
In particular loops with winding numbers greater than one will cancel against
products of loops with smaller winding numbers. To emphasize this we can rewrite
(\ref{zexpa}) as follows. We say that a loop on the new lattice corresponds
to a loop on the modified lattice if the same points can be traversed in the
same order. The amplitude of a loop on the new lattice corresponding to a
loop $r$ on the modified lattice (both with a winding number of one) shall be written as
$\tilde{L}_{r,k}$, where the index $k$ enumerates all such loops.
We can then write
\begin{equation}
L_{r}=\sum_{k=1}^{\infty}\tilde{L}_{r,k}
\end{equation}
Inserting this in (\ref{zexpa}) then gives
\begin{equation}\label{zxpn}
\frac{Z\haak{\ahaak{\epsilon}}}{Z_{0}}=\prod_{r=1}^{\infty}\rhaak{1+
\sum_{k=1}^{\infty}\tilde{L}_{r,k}}
\end{equation}

We shall
use the following notations for the bonds in a city. For a city at position $i$,
there is a bond with a weight proportional to $\epsilon_{i_{1}}^{k}$ and $\epsilon_{i_{2}}^{k}$,
if $k\geq 1$. These two bonds will be referred to as bonds of order $k$. A bond
that connects the points $a_{i}$ and $c_{i}$ will be referred to as a bond of
type 1, a bond that connects the points $b_{i}$ and $d_{i}$ will be referred to
as a bond of type 2. A diagram will be called balanced if at every city the sum 
of the orders of bonds of type 1 in the diagram equals the sum of the orders 
of bonds of type 2. From
(\ref{zpex}) it follows that only balanced diagrams contribute to $Z_{i_{1}\cdots i_{n}}$.
It is also clear that a particular balanced diagram contributes to $Z_{i_{1}\cdots i_{n}}$
only if at each city the sum of the orders of bonds of each type equals the
number of times the city occurs as an index in $Z_{i_{1}\cdots i_{n}}$. For 
balanced diagrams the sum of the orders of bonds of a certain type at a city
will be referred to as the order of the diagram at that city. If a city is not
part of a diagram, then the order of that diagram at that city is zero.
From (\ref{gzi}) it now follows that $g\haak{i_{1}\cdots i_{n}}$ can be written
as the sum of amplitudes of balanced diagrams, if the weights associated with
bonds are redefined as follows. We define the weights associated with the modified 
bonds, by choosing $\epsilon=\sqrt{U}$, so that a bond of order $k$ has a weight 
of $\pm\sqrt{2}\haak{\sqrt{U}}^{k}$. $g\haak{i_{1}\cdots i_{n}}$ can thus be 
written as a restricted sum of amplitudes of balanced diagrams, where the restriction
implies that a diagram contributes if at each city the order of the diagram equals 
the number of times the city occurs as an argument of $g$.
(\ref{zexg}) expresses $Z\haak{J=1}$ as
\begin{equation}
Z\haak{J=1}=Z\haak{J=0}\sum_{k=0}^{\infty}\frac{1}{k!}\sum_{i_{1}\cdots i_{k}}
g\haak{i_{1}\cdots i_{k}}
\end{equation}
By substituting the diagrammatic expansion of $g$ in this equation, we see that 
$\frac{Z\haak{J=1}}{Z\haak{J=0}}$ is a sum of balanced diagrams. We denote the
order of a balanced diagram at a city $j$ as $m_{j}$. We now focus on a balanced
diagram with $\sum_{j}m_{j}=k$. This diagram will appear $\frac{k!}{\prod_{j}m_{j}!}$
times in (\ref{zexg}). The contribution this diagram makes to $\frac{Z\haak{J=1}}{Z\haak{J=0}}$
is thus given by its amplitude multiplied by the factor $\frac{1}{\prod_{j}m_{j}!}$.
The fact that this extra factor appears is rather inconvenient. This problem
can be overcome by again changing the weights associated with the bonds. Since
diagrams which use more than one bond of a certain type at a city cancel, we can
absorb a factor $\frac{1}{\sqrt{r!}}$ in the weight of a bond of order $r$. The 
weight of a bond of order $r$ now becomes $\pm\sqrt{2}\frac{\haak{\sqrt{U}}^{r}}{\sqrt{r!}}$.
With this definition of the weights of the bonds $\frac{Z\haak{J=1}}{Z\haak{J=0}}$
is thus simply the sum of amplitudes of all balanced diagrams. Note that the factor
$\frac{1}{\sqrt{r!}}$ is rather arbitrary, we could have chosen $\frac{1}{\haak{r!}^{p}}$
for the weight of a bond of order $r$ of one type and $\frac{1}{\haak{r!}^{q}}$
for the bond of order r of the other type, provided that $p+q=1$. This arbitrariness
shall lead to a drastic simplification for the expansion of the free energy. 

We now proceed to write $\frac{Z\haak{J=1}}{Z\haak{J=0}}$ in a way analogous to
the expression (\ref{zexpn}) for $\frac{Z\haak{\ahaak{\epsilon}}}{Z_{0}}$, so 
that it becomes a simple matter to find the diagrammatic expansion of $\ln\haak{Z\haak{J=1}}$.
We can do this as follows. We define a diagram to be connected (irreducible), 
if it is balanced and cannot be written as a product of smaller balanced diagrams (by a smaller 
diagram we mean a diagram that contains only a part of the loops of the original
diagram). Suppose all connected diagrams are enumerated in some arbitrary order.
The amplitude of the $r^{\mbox{th}}$ connected diagram is denoted as $C_{r}$.
We can now express $\frac{Z\haak{J=1}}{Z\haak{J=0}}$ in terms of the $C_{r}$ as
\begin{equation}
\frac{Z\haak{J=1}}{Z_{0}}=\prod_{r=1}^{\infty}\rhaak{1+C_{r}}
\end{equation}
In this equation the product is over connected diagrams consisting of loops with 
winding numbers of one. The reduced free energy of the \stfm\ can thus be written as
\begin{equation}
\tilde{F}=F_{0}+\sum_{r=1}^{\infty}\ln\haak{1+C_{r}}
\end{equation}
As noted earlier, there is some freedom in the choice of the weights of the bonds.
At a certain city, only the product of the weights of the two bonds of order $r$
is required to contain the factor $\frac{1}{r!}$. For the expansion of the free energy
this means that a diagram that uses a bond of order $r$ of one type also has to
contain the bond of order $r$ of the other type, otherwise the diagram will cancel against
other diagrams. We can thus go back to the modified lattice by summing over the
orders of the bonds. The weight of a modified bond then becomes $\pm\sqrt{2\haak{\exp\haak{u}-1}}$.
The sign is given according to the orientation of the bond. The extra minus
sign can now be omitted, because any diagram has an even number of bonds.

\appendix
\chapter{Notations and conventions}
\section{Vectors}
Vectors are denoted like ordinary variables (i.e.\ without an arrow).
The modulus of a vector $ x $ is denoted as $\lhaak{x}$. By $ x^{n}$ with $n$ an even integer
we mean $\lhaak{x}^{n}$.
\section{Summation convention}
Repeated indices are summed over, unless stated otherwise.
\section{Multi-indices}\label{dfm}
A multi-index containing $ n $ indices is denoted as $\haak{n}$. We use a
summation convention in the following way:
\begin{equation}\label{scm}
A_{\haak{n}}B_{\haak{n}}=A_{i_{1}\ldots i_{n}}B_{i_{1}\ldots i_{n}}
\end{equation}
A possible sum over $n$ is always explicitly denoted. Different variables between brackets
always contain different indices. e.g.
\begin{equation}
A_{\haak{n}}B_{\haak{m}}=A_{i_{1}\ldots i_{n}}B_{j_{1}\ldots j_{m}}
\end{equation}
Identical variables between brackets contain identical indices (as in (\ref{scm})).
A tensor can have more than one multi-index:
\begin{equation}
A_{\haak{n},\haak{m}}=A_{i_{1}\ldots i_{n},j_{1}\ldots j_{m}}
\end{equation}
By 
\begin{equation}\sum_{\mbox{contractions}}A_{\haak{n_{1}},\haak{n_{2}}\ldots\haak{n_{r}}}\end{equation} 
we mean a sum over all possible contractions of the indices with each other,
e.g.
\begin{equation}\sum_{\mbox{contractions}}A_{\haak{n_{1}},\haak{n_{2}}}\end{equation}
with $ n_{1}=n_{2}=2 $ stands for
\begin{equation}A_{iikk}+A_{ikik}+A_{ikki}\end{equation}
\section{Derivatives}
For derivatives we sometimes use the notation $\partial^{\haak{a}}_{\haak{k},b}$. By
this we mean a the $ k^{\mbox{th}}$ derivative with respect to the components 
$\haak{k}$ of $b$ evaluated at the point $a$. The arguments $a$ and $b$ are 
optional.
\section{Distributions}\label{dfd}
For a detailed account of distributions we refer to \cite{rudin,swrt}.
Once a well defined space of functions $ S $ is defined, a space of continuous
linear functionals $ S'$ on $ S $ can be defined. We shall define the set $ S $
as the set of all functions $ h(x) $ that can be written as in (\ref{sg14}).
Apart from the usual $ L^{2} $-norm we define a Fr\'{e}chet norm as follows:
\begin{equation}
\lhaak{h}=\sup_{\haak{k},x,k\leq N}\lhaak{\partial_{\haak{k}}h(x)}
\end{equation}
Here $ N $ is some arbitrary number. The set of all continuous linear 
functionals (distributions) with respect to this norm is denoted by 
$ S_{N}' $. The action of a distribution $ T\in S_{N}' $ on $ h\in S $ is 
denoted as $ Th $ or as $ T(h) $. It is not difficult to see that the functional $ T $
defined as                                                
\begin{equation}Th=\lhaakr{\partial_{\haak{n}}h}_{z}\end{equation}
for some multi-index $\haak{n}$ is an element of $ S_{N}'$ when $ n\leq N $.
The functional $ I $ defined as
\begin{equation}Ih=\int I(x)h(x) d^{2}x\end{equation}
also defines a distribution as long as $ I(x) $ is bounded.
A weak-topology on $ S_{N}'$ is defined by saying that $ T_{n}\rightarrow T $
if $ T_{n}h\rightarrow Th\:\forall h\in S $. 
\chapter{Gaussian correlations}
\section{Correlation function for $\htw $}\label{dcor}
In this appendix we derive the expression (\ref{hr1}) for the correlation
function for the field $\htw $. By definition we have:
\begin{equation}\label{gx}
G\haak{x}=\gem{\htw\haak{x}\htw\haak{0}}
\end{equation}
In terms of $\hat{h}$ this becomes
\begin{equation}
G\haak{x}=\frac{1}{V}\sum_{k_{1},k_{2}}\gem{h\haak{k_{1}}h\haak{k_{2}}}e^{\imath k_{1}\cdot x}
\end{equation}
where the sum is over elements of the set $ S^{\haak{2}}$ (see section \ref{dfrn} for
the definition of this set). Since in the 
Gaussian model $ \hat{h}\haak{k_{1}} $ and $ \hat{h}\haak{k_{2}} $ are uncorrelated unless
$ k_{1}=\pm k_{2}$, we have
\begin{equation}\label{gx2}
G\haak{x}=\frac{1}{V}\sum_{k\in S^{\haak{2}}}\gem{\lhaak{\hat{h}\haak{k}}^{2}+\haak{\hat{h}\haak{k}}^{2}}e^{\imath k\cdot x}
\end{equation}
Now define real
valued variables $ a\haak{k} $ and $ b\haak{k} $ as follows:
\begin{equation}\label{adf}
h\haak{k}=a\haak{k}+\imath b\haak{k}
\end{equation}
Since $ h\haak{x} $ is real we have
\begin{equation}\label{bdf}
h\haak{-k}=a\haak{k}-\imath b\haak{k}
\end{equation}
From (\ref{adf}) and (\ref{bdf}) it follows that
\begin{equation}
dh\haak{k}dh\haak{-k}\equiv 2da\haak{k}db\haak{k}
\end{equation}
The partition function of the Gaussian model can be written as
\begin{equation}\label{part}
Z={\prod_{k\in S^{\haak{2}}}}'\int_{-\infty}^{\infty}\int_{-\infty}^{\infty}da\haak{k}db\haak{k}N\haak{k}
e^{-j\haak{k}k^{2}\haak{a\haak{k}^{2}+b\haak{k}^{2}}}={\prod_{k\in S^{\haak{2}}}}'\frac{\pi N\haak{k}}{j\haak{k}k^{2}}
\end{equation}
The prime indicates the condition $ k_y>0 $ and we have replaced the Gaussian coupling $ j $
by a $k$-dependent coupling $ j\haak{k} $. $ N\haak{k}$ is an uninteresting function 
coming from the definition of the measure $ Dh $ (see (\ref{maat})), and the 
Jacobian of the transformation to the variables $ a\haak{k}$ and $ b\haak{k}$.
It is clear from (\ref{part}) that
$\gem{\haak{\hat{h}\haak{k}}^{2}}$ vanishes. $\gem{\lhaak{\hat{h}\haak{k}}^{2}}$ is computed as follows:  
\begin{equation}\label{gx3}
\gem{\lhaak{\hat{h}\haak{k}}^{2}}=-\frac{1}{k^{2}}\lhaakr{\frac{\partial\ln\haak{Z}}{\partial{j\haak{k}}}}_{j\haak{k}=j}
=\frac{1}{jk^{2}}
\end{equation}
Using this one can write (\ref{gx2}) as: 
\begin{equation}\label{gx4}
G\haak{x}=\frac{1}{V}\sum_{k\in S^{\haak{2}}}\frac{e^{\imath k\cdot x}}{jk^{2}}
\end{equation}
We can formally rewrite (\ref{gx4}) using the characteristic function $ P_{c}^{\haak{2}}$ 
of the set $S^{\haak{2}}$ as:
\begin{equation}\label{gx5}
G\haak{x}=\int\frac{d^{2}k}{\haak{2\pi}^{2}}\frac{P^{(2)}\haak{k}}{jk^{2}}
e^{\imath k\cdot x}
\end{equation}
\section{The height-height correlation function}\label{hhcr}
In this section we will evaluate the height-height correlation function ($ R(x) $)
for the Gaussian model. By definition we have
\begin{equation}
R\haak{x}=\gem{\haak{h\haak{x}-h\haak{0}}^{2}}
\end{equation}
Using (\ref{hr1}) we can write this as
\begin{equation}
R\haak{x}=-\frac{2}{j}\int\frac{d^{2}k}{\haak{2\pi}^{2}}\frac{e^{\imath k\cdot x}-1}{k^{2}}P\haak{k}
\end{equation}
Integrating over the angle between $k$ and $x$ yields
\begin{equation}
R\haak{x}=-\frac{1}{\pi j}\int_{0}^{\infty}\frac{d\lhaak{k}}{\lhaak{k}}\haak{J_{0}\haak{\lhaak{k}\lhaak{x}}-1}P\haak{k}
\end{equation}
Differentiating both sides of this equation with respect to $\lhaak{x}$ yields
\begin{equation}
R'\haak{x}=-\frac{1}{\pi j \lhaak{x}}\int_{0}^{\infty}d\lhaak{k}J_{0}'\haak{\lhaak{k}\lhaak{x}}P\haak{k}
=-\frac{1}{\pi j \lhaak{x}}\int_{0}^{\infty}dt J_{0}'\haak{t}P\haak{\frac{t}{\lhaak{x}}}
\end{equation}
where we have substituted $t=\lhaak{k}\lhaak{x}$.
For large $x$ the integral is $-1$. We thus have
\begin{equation}
R\haak{x}\sim\frac{1}{\pi j}\ln\haak{\lhaak{x}}
\end{equation}
\chapter{Calculation of $\gem{e^{\imath T h }}$}\label{dfund2}
\noindent In this appendix we derive equation (\ref{fund2}). First we derive this 
equation in the special case of a distribution $ T $ that can be written
as a finite linear combination of Dirac delta's. By taking appropriate
limits of this special case we arrive at (\ref{fund2}).

Let the distribution $ T $ be defined by:
\begin{equation}\label{tdf}
Th=\sum_{n}\alpha_{n}h\haak{x_{n}}
\end{equation}
In this case we have
\begin{equation}
\gem{\exp\haak{\imath Th}}=\exp\haak{\imath T\hee}\gem{\exp\haak{\imath T\htw}}
\end{equation}
Using (\ref{sg16}) we get 
\begin{equation}
\begin{array}{ll}
\gem{\exp\haak{\imath T\htw}}&=\int D\htw \exp\haak{\imath \sum_{n}\alpha_{n}\htw\haak{x_{n}}}\\
&\times\exp\haak{-\frac{j}{2}\int\haak{\nabla\htw}^{2}d^{2}x}\\
\end{array}
\end{equation}
In terms of $ h\haak{k} $ this can be written as
\begin{equation}\label{ap1}
\int D\htw \exp\haak{-\sum_{k}\rhaak{\frac{j}{2}k^{2}\lhaak{h\haak{k}}^{2}-\imath\sum_{n}\frac{1}{\sqrt{V}}
h\haak{k}\exp\haak{\imath k\cdot x_{n}}}}
\end{equation}
Here the sum is over those $k$-values that belong to $ \htw $. 
We write $ h\haak{k}=a\haak{k}+\imath b\haak{k} $, with $ a\haak{k} $ and $ b\haak{k} $ real.
The integral (\ref{ap1}) can be written as:
\begin{equation}\label{ap3}
\begin{array}{l}
\rhaakl{{\prod_{k}}'\int 2da\haak{k}db\haak{k}\exp\haakl{-jk^{2}\haak{a\haak{k}^{2}+b\haak{k}^{2}}}}\\
\rhaakr{\haakr{+\frac{2\imath}{\sqrt{V}}\sum_{n}\rhaak{\alpha_{n}\haak{a\haak{k}\cos\haak{k\cdot x_{n}}
-b\haak{k}\sin\haak{k\cdot x_{n}}}}}}\\
\times\rhaak{{\prod_{k}}'\int 2da\haak{k}db\haak{k}\exp\haak{-jk^{2}\haak{a\haak{k}^{2}+b\haak{k}^{2}}}}^{-1}\\
\end{array}
\end{equation}
Here the prime indicates that the product is over $k$-values with $ k_{y}>0 $.
Using 
\begin{equation}
\int_{-\infty}^{\infty}\exp\haak{-ax^{2}+bx}dx=\exp\haak{\frac{b^{2}}{4a}}\sqrt{\frac{\pi}{a}}
\end{equation}
the integrals over $ a\haak{k} $ and $ b\haak{k} $ are easily performed. (\ref{ap3}) simplifies to
\begin{equation}\label{app4}
{\prod_{k}}'\exp\haak{-\frac{1}{Vjk^{2}}\rhaak{\haak{\sum_{n}\alpha_{n}\cos\haak{k\cdot x_{n}}}^{2}
+\haak{\sum_{n}\alpha_{n}\sin\haak{k\cdot x_{n}}}^{2}}}
\end{equation}
The product is evaluated by writing it as an exponentiated sum, and then replacing
the sum by an integral. We thus have 
\begin{equation}\label{app5}
\begin{array}{l}
{\prod_{k}}'\exp\haak{-\frac{1}{Vjk^{2}}\rhaak{\haak{\sum_{n}\alpha_{n}\cos\haak{k\cdot x_{n}}}^{2}
+\haak{\sum_{n}\alpha_{n}\sin\haak{k\cdot x_{n}}}^{2}}}\\
=\exp\haak{-{\sum_{k}}'\frac{1}{Vjk^{2}}\sum_{n,m}\alpha_{n}\alpha_{m}\cos\haak{k\cdot\haak{x_{n}-x_{m}}}}\\
\end{array}
\end{equation}
Since the density of degrees of freedom for the field $ h^{2} $ is $ P^{\haak{2}}\haak{k}$
(see (\ref{sg7})) we can replace a sum by an integral as follows:
\begin{equation}
\sum_{k}\rightarrow V\int_{k}\frac{d^{2}k}{\haak{2\pi}^{2}}P^{\haak{2}}\haak{k}
\end{equation}
We can thus write the exponent of (\ref{app5}) as
\begin{equation}\label{app6}
-\frac{1}{2j}\int\frac{d^{2}k}{\haak{2\pi}^{2}}\frac{1}{k^{2}}P^{\haak{2}}\haak{k}
\sum_{n,m}\alpha_{n}\alpha_{m}\cos\haak{k\cdot\haak{x_{n}-x_{m}}}
\end{equation}
The extra factor $\frac{1}{2}$ comes from taking into account the condition $ k_{y}>0 $.  
Using (\ref{hr1}) we can write this as
\begin{equation}
-\frac{1}{2}\sum_{n,m}\alpha_{n}\alpha_{m}G\haak{x_{n}-x_{m}}
\end{equation}
from which it follows that
\begin{equation}\label{appr1}
\begin{array}{l}
\gem{\exp\haak{\imath\sum_{n}\alpha_{n}h\haak{x_{n}}}}\\
=\exp\haak{\imath\sum_{n}\alpha_{n}\hee\haak{x_{n}}}
\exp\haak{-\frac{1}{2}\sum_{n,m}\alpha_{n}\alpha_{m}G\haak{x_{n}-x_{m}}}\\
\end{array}
\end{equation}
This proves (\ref{fund2}) in this special case. Proving the general case is 
a matter of taking limits in (\ref{appr1}). First we take the continuum limit
in (\ref{appr1}). Let $ T $ be a distribution generated by a bounded continuous function 
$ T(x) $ i.e.\ $ Th=\int d^{2}\! x T(x)h(x) $. Then since there exist distributions $ T_{n} $
of the form (\ref{tdf}), such that
\begin{equation}\lim_{n\rightarrow\infty}T_{n}h=Th\end{equation}
(\ref{fund2}) is also valid in this case. Since for every distribution $ u $ there
exist continuous functions $ u_{n}(x) $ such that 
\begin{equation}\lim_{n\rightarrow\infty}u_{n}h=uh\end{equation}
(see \cite{rudin}), (\ref{fund2}) follows.

\end{document}